\documentstyle[tighten,aps,axodraw,float]{revtex}

\newcommand{\be}{\begin{equation}}
\newcommand{\ee}{\end{equation}}
\newcommand{\ba}{\begin{eqnarray}}
\newcommand{\ea}{\end{eqnarray}}

\newcommand{\ep}{\varepsilon}

\def\slash#1{\setbox0=\hbox{$#1$}               
   \dimen0=\wd0                                 
   \setbox1=\hbox{/} \dimen1=\wd1               
   \ifdim\dimen0>\dimen1                        
      \rlap{\hbox to \dimen0{\hfil/\hfil}}      
      #1                                        
   \else                                        
      \rlap{\hbox to \dimen1{\hfil$#1$\hfil}}   
      /                                         
   \fi}                                         %

\begin{document}

\title{
%
%
\[ \vspace{-2cm} \]
\noindent\hfill\hbox{\rm  SLAC-PUB-8450, TTP00-08} \vskip 1pt
\noindent\hfill\hbox{\rm hep-ph/0005131} \vskip 10pt
%
%
The three-loop on-shell renormalization of QCD and QED }

\author{Kirill Melnikov\thanks{
e-mail:  melnikov@slac.stanford.edu}}
\address{Stanford Linear Accelerator Center\\
Stanford University, Stanford, CA 94309}
\author{Timo van Ritbergen\thanks{
e-mail:  timo@particle.physik.uni-karlsruhe.de}}
\address{Institut f\"{u}r Theoretische Teilchenphysik,\\
Universit\"{a}t Karlsruhe,
D--76128 Karlsruhe, Germany}
\maketitle

\begin{abstract}
We describe a calculation of the {\it on-shell} 
renormalization factors in QCD and QED at the three loop
level. Explicit results for the fermion  
mass renormalization factor $Z_m$ and the on-shell
fermion  wave function renormalization constant $Z_2$ are given. 
We find that at ${\cal O}(\alpha_s^3)$ the wave function renormalization 
constant $Z_2$ in QCD becomes gauge dependent also in the on-shell scheme, 
thereby disproving the ``gauge-independence'' conjecture 
based on an earlier two-loop result.
As a byproduct, we derive an ${\cal O}(\alpha_s^3)$ contribution 
to the anomalous dimension of the heavy quark field in HQET.
\end{abstract}


\section{Introduction}
Perturbative calculations in field theory are often instrumental 
 in establishing accurate relations between theory and experiment. 
For example, in heavy quark physics, that now occupies the central place 
in the study of ${\rm CP}$ violation, perturbative effects play an important 
role alongside with non-perturbative effects.
For heavy quarks, one separates  soft and hard contributions
through an  expansion in the inverse quark 
mass; familiar examples being HQET (for a review see e.g.\cite{neubert})
and NRQCD \cite{bbl}.
It often happens that the leading term in such an expansion is determined by 
a perturbative calculation in full QCD and  for this reason   
perturbative calculations with heavy {\it on-shell}
quarks in the initial and final state received significant attention 
in recent years. For many processes of interest we have 
witnessed   a rapid computational
progress  and currently the ``standard'' level of accuracy 
is the next-to-next-to-leading or ${\cal O}(\alpha_s^2)$ order. 
Further progress in this direction 
will require three-loop computations.

In QED, an interesting recent development was an application of 
dimensionally regularized NRQED to the calculation of fundamental
properties of simple atoms, like positronium and hydrogen.
Here too, higher precision will require on-shell 
three loop calculations in many cases.

The renormalizability of QCD and QED implies that  
any multi-loop calculation can only be made meaningful 
if the corresponding renormalization constants are known\footnote{An
immediate consequence of a more precise knowledge of just the
  on-shell renormalization factors is that it leads to an exact 3-loop result
  for the relation between the pole-quark-mass and
   the ${\overline {\rm MS}}$-quark-mass, which was recently presented 
    in \cite{uns}.}; 
hence the three-loop computations require 
the knowledge of the three-loop renormalization constants in QCD.
As is well known, the renormalization constants are 
renormalization scheme dependent.
In some  schemes, as e.g. in the $\overline {\rm MS}$ scheme, they have been 
calculated up to the fourth order in perturbation theory. 
This is however insufficient
if one is interested in the processes with heavy quarks being on-shell
in the initial and final state -- precisely as in QED, the renormalization 
constants in the on-shell scheme are required.

If one adopts the on-shell renormalization scheme for the quarks in QCD, 
two specific renormalization constants need to be calculated. 
The mass renormalization constant $Z_m$ 
is defined by $m_0 = Z_m m$, where $m_0$ is the bare and $m$ is
the pole mass. The wave function renormalization factor $Z_2$ is defined as 
$\psi_0 = \sqrt{Z_2} \psi$, where $\psi_0$ and
$\psi$ are the bare and renormalized quark fields respectively.
Both renormalization constants can be obtained from the expression 
for the heavy quark propagator close to the mass shell. The 
mass renormalization constant $Z_m$ is determined from the position 
of the pole of the heavy quark propagator, while $Z_2$ is determined
from the residue.  Because of the infrared catastrophe, the residue
does not exist  and the introduction of the 
infrared regulator becomes necessary. In what follows   
dimensional regularization is used to regulate both ultraviolet 
{\it and} infrared divergences  which appear in the calculation. 

Let us remark that the choice of the infrared regulator is 
not important for the pole mass, since it is known  
to be infrared finite and  gauge invariant to all orders in perturbation  
theory.  On the contrary, $Z_2$ is not infrared finite already 
in the first non-trivial order in perturbative expansion.
Hence, choosing the infrared regulator merely amounts to 
defining $Z_2$ (in addition to  usual scheme definition) 
and one may expect that the question of gauge 
invariance may  then depend on that choice\footnote{In what follows, we 
loosely use the terms ``gauge invariant'', ``gauge independent'' and 
``gauge parameter independent'' as equivalent.}.
Indeed, that is exactly what happens. An interesting phenomenon
occurs if dimensional regularization is applied to infrared
divergences.  In this situation,  in QED, $Z_2$ becomes gauge invariant to all
orders in the coupling constant. This feature follows from dimensionally
regulated Johnson-Zumino identity \cite{landau,johnson,fukuda}:
\be
\frac {{\rm d} \log Z_2 }{{\rm d} \xi} = -ie_0^2 
\int \frac {{\rm d}^D k}{(2\pi)^D k^4} = 0,
\label{eq1}
\ee
where $\xi$ is the gauge fixing parameter in a general covariant gauge.
Note that the right hand side of the above equation is zero 
by virtue of the fact that scale-less integrals in dimensional 
regularization are {\it defined} to be zero. We will derive Eq.(\ref{eq1})
in Section VI using  the path integral formalism.

Though no relation similar to Eq.(\ref{eq1}) 
has been derived in QCD, the use of 
dimensional regularization for infra-red divergences resulted in  
some interesting observations.  The one-loop $Z_2$ in QCD is gauge 
invariant since it is a trivial generalization of the QED result.
The two-loop contribution to $Z_2$ is already sensitive to the 
non-abelian structure of QCD and there is no a priori reason 
to expect it to be gauge invariant. However, an explicit computation 
in  Ref.\cite{broadhurst} produced  a gauge invariant result.
The authors of that reference then conjectured that this  fact
may be true to all orders in perturbation theory, 
though no supporting arguments have been given.  

The purpose of this paper is to present the three loop calculation 
of $Z_m$ and $Z_2$ in dimensional regularization and hence to
give a complete  set of  three loop renormalization constants 
that are specific to the on-shell scheme.  
The knowledge of $Z_m$ and $Z_2$ with 
such an accuracy is an important step towards three loop calculations 
with heavy quarks close to the mass shell. The bulk of the paper is technical 
and attempts to provide some details about how the actual
computation is done. Though the scale of this calculation
makes it impossible to provide a detailed
account of the algorithms used to obtain the result,
we at least try to highlight some of its aspects
which might be of importance for future related work.
A peculiar feature of our result is that, in contrast to two first orders 
in perturbative expansion,  the ${\cal O}(\alpha_s^3)$ contribution 
to $Z_2$ turns  out to be {\it gauge dependent} thus disproving 
the all-orders gauge independence  conjecture of Ref.\cite{broadhurst}.

The paper is organized as follows. In the next Section we introduce 
the necessary notations and describe the calculation. In Section III 
we discuss in detail how new master integrals that appear in non-abelian
theory  are computed. In Section IV the results for QCD renormalization 
constants are summarized.  In Section V  the anomalous dimension 
of the heavy quark field in HQET is derived. In Section VI
we give the three loop renormalization  constants 
for QED. In the appendices a complete list of master 
integrals for the three loop on-shell two-point functions, 
as well as some other useful formulas,  are presented.

\section{Preliminaries}

We are going to calculate the heavy quark mass and the heavy 
quark wave function renormalization constants in the on-shell
renormalization scheme.  The bare quark mass $m_0$ and the 
bare quark field $\psi_0$  are renormalized multiplicatively:
\be
m_0 = Z_m m,~~~~~~~~~~~~~~\psi_0 = \sqrt{Z_2} \psi,
\ee
where $m$ and $\psi$ stand for the pole mass and the renormalized
quark field, respectively. Both renormalization constants can be 
derived by considering  one-particle irreducible quark self-energy 
operator $\hat \Sigma(p,m)$. We discuss this derivation in some detail 
since we believe that  discussion given in \cite{broadhurst} 
is  slightly over-complicated  in that place.

Because of the  Lorentz invariance, 
the one-particle irreducible quark self energy 
operator $\hat \Sigma(p,m)$   can be parameterized by
two independent functions:
\be
\hat \Sigma(p,m) = m \Sigma_1(p^2,m) + (\hat p - m)\Sigma_2(p^2,m),
\ee
so that complete fermion propagator reads:
\be
\hat S_F(p) = \frac {i}{\hat p - m_0 + \hat \Sigma(p,m)}.
\label{fermprop}
\ee
Let us expand the self energy operator in  formal Taylor series
around $p^2=m^2$:
\ba
 \hat \Sigma(p,m) &&\approx  m \left. \Sigma_1(p^2,m)\right|_{p^2=m^2} +
   m \frac {\partial}{\partial p^2} 
 \left. \Sigma_1(p^2,m)\right|_{p^2=m^2} (p^2-m^2) +
   (\hat p - m) \left. \Sigma_2(p^2,m)\right|_{p^2=m^2}
\nonumber \\
&&\approx  m \Sigma_1(p^2,m)|_{p^2=m^2} + 
  (\hat p - m) \left ( 2m^2  \frac {\partial}{\partial p^2} 
 \left. \Sigma_1(p^2,m)\right|_{p^2=m^2}
   + \left. \Sigma_2(p^2,m)\right|_{p^2,m^2}  \right ).
\ea                       
Substituting this expression into the fermion propagator Eq.(\ref{fermprop}),
we identify the position of the pole with the pole mass and the 
residue with the wave function renormalization constant.  
We  obtain:
\ba  \label{ZOSsigma}
&&Z_m = 1 + \left. \Sigma_1(p^2,m)\right|_{p^2=m^2},
\nonumber \\
&&\frac {1}{Z_2} = 1 +  2m^2 \frac {\partial}{\partial p^2} 
 \left. \Sigma_1(p^2,m)\right|_{p^2=m^2}
   + \left. \Sigma_2(p^2,m)\right|_{p^2=m^2}.
\label{e6}
\ea

Both of these constants can be obtained easily once an expansion 
of $\hat \Sigma(p,m)$ close to the mass shell 
is available. Let us introduce the 
Minkowski  vector $Q$ with $Q^2 = m^2$ and consider a one-parameter
family of external momenta $p = Q(1+t)$. The quark self-energy 
operator reads:
\be
\hat \Sigma(p,m) = m \Sigma_1(p^2,m) + (\hat Q - m)  \Sigma_2(p^2,m)
+ t \hat Q \Sigma_2(p^2,m).
\ee
Taking the trace
\be
T_1 = {\rm Tr} \left [ \frac{\left ( \hat Q + m \right ) }{4m^2} 
 \hat \Sigma(p,m) \right ] = 
\Sigma_1(p^2,m) + t\Sigma_2(p^2,m),
\label{trace}
\ee
and expanding in $t$ up to ${\cal O}(t)$ we obtain:
\be
T_1 = \left. \Sigma_1(p^2,m)\right|_{p^2=m^2} +t \left [
2m^2 \frac {\partial}{\partial p^2} 
 \left. \Sigma_1(p^2,m)\right|_{p^2=m^2}
+ \left. \Sigma_2(p^2,m) \right|_{p^2=m^2} \right ] + {\cal O}(t^2).
\label{e9}
\ee
Comparing Eqs.(\ref{e6}) with Eq.(\ref{e9}), 
we see  that expanding  the self energy operator in $Qt$ 
along with projecting on the relevant structure by taking the trace
as shown in Eq.(\ref{trace}) delivers an expression one 
needs to reconstruct the on-shell mass and  wave function renormalization 
constants. The pole mass in the above equation is calculated iteratively 
by using  mass counterterms where appropriate, i.e. by calculating lower 
 order diagrams with the appropriate mass counterterm insertion.

Hence, both $Z_m$ and $Z_2$ can be extracted from 
the heavy quark self-energy operator evaluated on-shell. 
This task is rather straightforward in low orders of perturbation 
theory but it becomes increasingly  difficult when one 
goes to higher orders. Both the number of diagrams one
has to calculate and the complexity of integrals increases
dramatically. In the present case we have about sixty diagrams
to be calculated and the question about an efficient 
way to do the calculation becomes of tremendous importance.

The most efficient way to evaluate those multiloop integrals 
is to utilize integration-by-parts identities within dimensional 
regularization \cite{thooft,ibp} . The first thing to realize is 
that  {\it any}  integral that one can face in computing the on-shell 
fermion propagator can be expressed through eleven 
basic three loop integrals (topologies) shown in 
Fig.1.  Any integral that belongs to a certain topology
is considered to be a function of the powers of denominators
(both positive and negative integer powers are allowed)
and one irreducible numerator in each case.
For each of the topologies one writes down a system of 
recurrence relations based on integration-by-parts identities 
that are derived using the fact 
that {\it in dimensional regularization}
any integral of the total derivative vanishes:
\be
0 = \int \frac {\partial }{\partial k_i^{\mu}} 
\left ( l_j^{\mu} \times {\rm propagators} \right ).
\ee
Here $k_i$  is one of the three loop momenta and 
$l_j$ is either one of the loop momenta or the  external 
momenta. Hence the starting set of equations consists
of twelve recurrence relations for each of the topologies.
The fact that not all of these relations are independent 
is not an obstacle since, when trying to solve them,
some relations will vanish identically and for this 
reason will automatically be of no use.

The next step is to solve the system of these equations.  Though 
there are several ways of thinking about what such  a solution 
might be, we prefer to look for the most general one. 
We then want to construct an algorithm
that, for a given topology and for any given initial set of 
powers of propagators, expresses an initial integral
through a minimal set of ``simpler'' integrals.
The simpler integrals are usually those that either have
denominators raised to small powers  or those that belong to simpler
topologies. We then consider these ``simpler'', but still 
non-trivial topologies, write down a new set of recurrence
relations for them, construct an algorithm that reduces
any integral to even simpler topologies and continue along
these lines until  we have an algorithm that 
completely solves the initial problem in terms of a few
master integrals. 
The final set of master integrals is found experimentally. 
There is no proof that the set we find is indeed minimal 
with respect to integration-by-parts relations
in the strict mathematical sense, but for practical calculations
this set of integrals is sufficient.

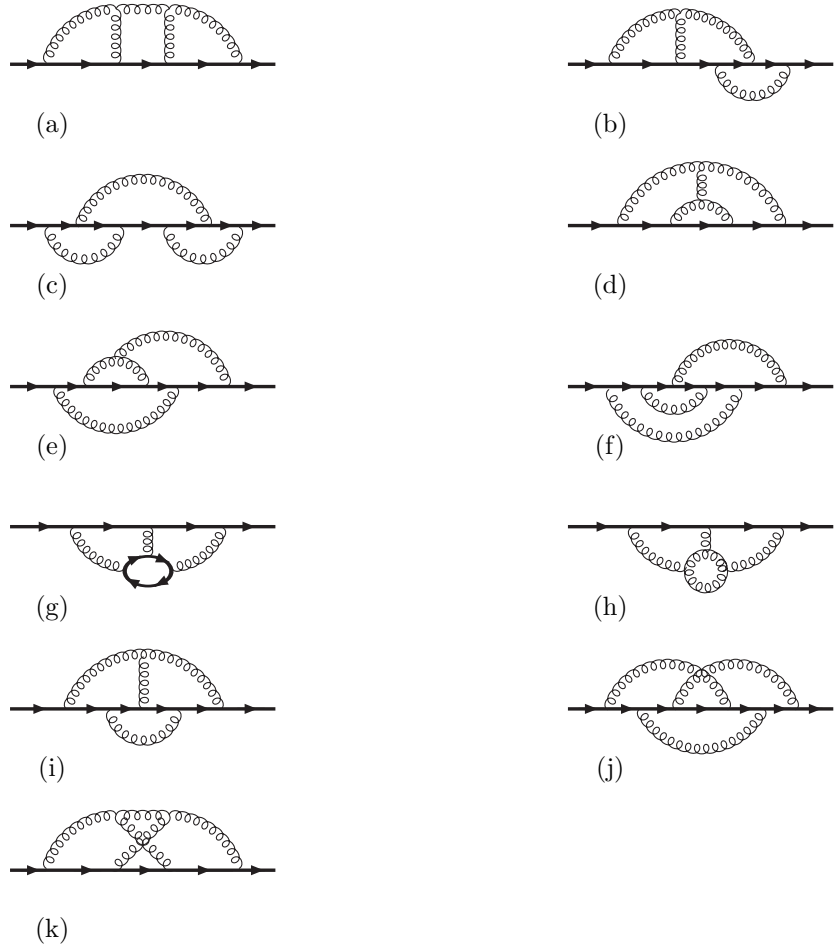
\begin{figure}
\begin{center}
\hfill
\begin{picture}(120,40)(0,0)
  \SetScale{.8}
 \SetWidth{1.6}
 \Line(5,20)(130,20)
 \SetWidth{1}
  \ArrowLine(10,20)(22,20)
 \ArrowLine(35,20)(48,20)
 \ArrowLine(60,20)(80,20) 
 \ArrowLine(83,20)(110,20) 
 \ArrowLine(118,20)(125,20)
 \SetWidth{0.5}
 \GlueArc(80,12)(34,15,90){2.2}{9.5}  
 \GlueArc(55,12)(34,90,165){2.2}{9.5}
 \Gluon(80,20)(80,46){2.2}{5}
 \Gluon(55,20)(55,46){-2.2}{5}
 \Gluon(80,46)(55,46){-2.2}{4}
 \Text(20,-2)[t]{(a)}
\end{picture}
\hfill
\begin{picture}(120,40)(0,0)
 \SetScale{.8}
 \SetWidth{1.6}
 \Line(5,20)(130,20)
 \SetWidth{1}
  \ArrowLine(10,20)(22,20)
 \ArrowLine(38,20)(48,20)
 \ArrowLine(60,20)(80,20)
 \ArrowLine(82,20)(90,20)
 \ArrowLine(92,20)(110,20)
 \ArrowLine(114,20)(125,20)
 \SetWidth{0.5}
 \GlueArc(58,8)(36,21,90){2.2}{10.5}
 \GlueArc(58,10)(34,90,162){2.2}{9.5}
 \Gluon(58,20)(58,44){2.2}{5}
 \GlueArc(92,21)(16,186,354){2.2}{9.5}
\Text(20,-2)[t]{(b)}
\end{picture}
\hfill\null\\
\hfill
\begin{picture}(120,60)(0,0)
 \SetScale{.8}
 \SetWidth{1.6}
 \Line(5,20)(130,20)
 \SetWidth{1}
 \ArrowLine(10,20)(22,20)
 \ArrowLine(24,20)(40,20)
 \ArrowLine(40,20)(55,20)
 \ArrowLine(60,20)(80,20)
 \ArrowLine(86,20)(96,20)
 \ArrowLine(104,20)(110,20)
 \ArrowLine(119,20)(125,20)
 \SetWidth{0.5}
 \GlueArc(96,21)(17,186,354){2.2}{9.5}
 \GlueArc(40,21)(17,186,354){2.2}{9.5}
 \GlueArc(68,10)(32,17,163){2.2}{17.5}
 \Text(20,-2)[t]{(c)}
\end{picture}
\hfill
\begin{picture}(120,60)(0,0)
  \SetScale{.8} 
 \SetWidth{1.6}
 \Line(5,20)(130,20)
 \SetWidth{1}
  \ArrowLine(16,20)(22,20)
 \ArrowLine(36,20)(50,20)
 \ArrowLine(60,20)(80,20)
 \ArrowLine(90,20)(100,20)  
 \ArrowLine(112,20)(125,20)
 \SetWidth{0.5}
 \GlueArc(68,16)(14,17,163){2.2}{7.5}
 \GlueArc(68,8)(40,17,163){2.2}{21.5}
  \Gluon(68,32)(68,46){2.2}{3.5}
 \Text(20,-2)[t]{(d)}
\SetScale{1}
\end{picture}
\hfill\null\\
\hfill
\begin{picture}(120,60)(0,0)
 \SetScale{.8}
 \SetWidth{1.6}
 \Line(5,20)(130,20)
 \SetWidth{1}
  \ArrowLine(10,20)(22,20)
 \ArrowLine(28,20)(42,20)
 \ArrowLine(50,20)(62,20)
 \ArrowLine(68,20)(87,20)
 \ArrowLine(94,20)(99,20)  
 \ArrowLine(112,20)(125,20)
 \SetWidth{0.5}
 \GlueArc(55,18)(14,10,170){2.2}{8.5}   
 \GlueArc(55,30)(30,200,340){2.2}{16.5}   
 \GlueArc(78,14)(30,12,140.5){2.2}{14}  
 \Text(20,-2)[t]{(e)}
\end{picture}
\hfill
\begin{picture}(120,60)(0,0)
  \SetScale{.8}
 \SetWidth{1.6}
 \Line(5,20)(130,20)
 \SetWidth{1}
  \ArrowLine(10,20)(22,20)
 \ArrowLine(25,20)(42,20)
 \ArrowLine(42,20)(58,20)  
 \ArrowLine(58,20)(68,20)
 \ArrowLine(68,20)(87,20)
 \ArrowLine(92,20)(100,20)
 \ArrowLine(112,20)(125,20)
 \SetWidth{0.5}
 \GlueArc(55,24)(15,197,343){2.2}{7.5}
 \GlueArc(55,28)(32,197,343){2.2}{16.5}
 \GlueArc(81,14)(26,13,167){2.2}{16.5}
 \Text(20,-2)[t]{(f)}
\end{picture}
\hfill\null\\
\hfill
\begin{picture}(120,60)(0,0)
    \SetScale{.8}
 \SetWidth{1.6}
 \Line(5,30)(130,30)
  \SetWidth{1.3}
  \Oval(70,9)(7,11)(0)
 \SetWidth{1}
 \ArrowLine(15,30)(28,30)
 \ArrowLine(42,30)(62,30)
 \ArrowLine(83,30)(99,30)
 \ArrowLine(115,30)(120,30)
  \ArrowLine(61.5,14)(66.5,16)
  \ArrowLine(74,15.5)(80,13)
   \ArrowLine(79,5.0)(73,2)
   \ArrowLine(65,2.7)(60,5.5)
 \SetWidth{0.5}
  \Gluon(70,30)(70,16){2.2}{3} 
 \GlueArc(59,36)(25,195,270){2.2}{7.5}
 \GlueArc(81,36)(25,270,345){2.2}{7.5}
 \Text(20,-2)[t]{(g)}
\end{picture}
\hfill
\begin{picture}(120,60)(0,0)
 \SetScale{.8}
 \SetWidth{1.6}
 \Line(5,30)(130,30)
  \SetWidth{1.3}
 \SetWidth{1}
 \ArrowLine(15,30)(28,30)
 \ArrowLine(42,30)(62,30)  
 \ArrowLine(83,30)(99,30)
 \ArrowLine(115,30)(120,30) 
 \SetWidth{0.5}
 \GlueArc(70,9)(8,0,400){2.2}{12.8}
 \Gluon(70,30)(70,18){2.2}{2.5}
 \GlueArc(59,36)(25,195,274){2.2}{7.5}
 \GlueArc(81,36)(25,266,345){2.2}{7.5}
 \Text(20,-2)[t]{(h)}
\end{picture}
\hfill\null\\
\hfill
\begin{picture}(120,60)(0,0)
 \SetScale{.8}
 \SetWidth{1.6}
 \Line(5,20)(130,20)
 \SetWidth{1}
  \ArrowLine(16,20)(22,20)
 \ArrowLine(38,20)(50,20)
 \ArrowLine(50,20)(70,20)
 \ArrowLine(74,20)(80,20)
 \ArrowLine(92,20)(100,20)
 \ArrowLine(112,20)(125,20)
 \SetWidth{0.5}
 \GlueArc(68,8)(38,18,162){2.2}{21.5}
  \Gluon(68,20)(68,44.5){2.2}{5.5}
  \GlueArc(68,21)(16,186,354){2.2}{9.5}  
 \Text(20,-2)[t]{(i)}
\end{picture}
\hfill
\begin{picture}(120,60)(0,0)
 \SetScale{.8} 
 \SetWidth{1.6}
 \Line(5,20)(130,20)
 \SetWidth{1}
 \ArrowLine(10,20)(22,20)
 \ArrowLine(24,20)(42,20) 
 \ArrowLine(42,20)(58,20) 
 \ArrowLine(59,20)(78,20)
 \ArrowLine(82,20)(96,20) 
 \ArrowLine(100,20)(110,20)
 \ArrowLine(118,20)(125,20)
 \SetWidth{0.5}
 \GlueArc(84,12)(29.2,15,165){2.2}{16.5}
 \GlueArc(68,33)(32,204,336){2.2}{16.5}
 \GlueArc(52,12)(29.2,15,165){2.2}{16.5}
\Text(20,-2)[t]{(j)}
\end{picture}
\hfill\null\\
\hfill
\begin{picture}(120,60)(0,0)
 \SetScale{.8}
\SetWidth{1.6}
 \Line(5,20)(130,20)
 \SetWidth{1}
  \ArrowLine(10,20)(22,20)
 \ArrowLine(35,20)(48,20)
 \ArrowLine(58,20)(80,20)
 \ArrowLine(83,20)(110,20)
 \ArrowLine(118,20)(125,20)
 \SetWidth{0.5}
 \GlueArc(80,12)(34,15,90){2.2}{9.5}
 \GlueArc(55,12)(34,90,165){2.2}{9.5}
  \Gluon(82,20)(55,46){2.2}{7.5}
  \Gluon(53,20)(80,46){-2.2}{7.5}
 \Gluon(80,46)(55,46){-2.2}{4}
 \Text(20,-2)[t]{(k)}
\end{picture}
\hfill
\begin{picture}(120,60)(0,0)
\end{picture}
\hfill\null\\
\vglue 18pt
\end{center}
\caption{Examples of three-loop quark propagator diagrams corresponding to
   eleven integration topologies.}
\label{fig:examplediagrams}
\end{figure}

Our solution of the system of recurrence
relations shows that  it is possible to express any integral which 
belongs to the above topologies through $18$ 
master integrals. Most of these integrals have  been 
calculated in the course of the analytical calculation 
of the electron anomalous magnetic moment 
\cite{Laporta} and can be taken from there.
It is remarkable that a transition from the abelian theory to the non-abelian
theory does not result in a significant increase in the number of 
master integrals to be computed, although the number of basic topologies
does.  As compared to Ref.\cite{Laporta}, we need one additional
master integral that corresponds to topology A and we also 
need one of the master integrals
of Ref. \cite{Laporta} to a higher order in the regularization
parameter $\varepsilon$. For the QCD wave function renormalization
constant we also need the constant $C_1$ (see \cite{Laporta} ) 
which was not computed  in \cite{Laporta}, because it 
mysteriously canceled in the calculation of the electron anomalous
magnetic moment\footnote{Moreover, we observed a similar cancellation 
in our calculation of the three loop slope of 
the Dirac form factor \cite{chrad} as well as 
in the calculation of the relation between the $\overline {\rm MS}$ and
the pole quark masses \cite{uns}.}. The calculation of these new
master integrals  is described in the next Section.

There are two principal checks on our solution of the recurrence relations.
First, we have computed the three-loop anomalous magnetic moment of the 
 electron and confirmed the result of Ref.\cite{Laporta}. Second, the actual
calculation of the on-shell  quark mass renormalization constant 
$Z_m$ has been performed in  an arbitrary covariant gauge 
for the gluon field. The explicit
cancellation of the gauge parameter in our result for $Z_m$
is an important check of the correctness of the calculation.
There are also some checks related to anomalous dimensions 
of the heavy quark field  in HQET, that can be computed 
combining our result for 
$Z_2$ with the known result for $Z_2^{\overline {\rm MS}}$.
We will elaborate on this issue in Section V.

\section{New master integrals}

In this Section the calculation of three new master integrals is described.
Note that throughout this paper we work with the integrals defined in 
Euclidean space. A pictorial representation of all master integrals 
is given in Fig.2 in the appendix.
 The space-time dimension in what follows is parameterized
by $D=4-2\ep$. We will also use  $C(\ep) = [\pi^{2-\ep} \Gamma(1+\ep)]^3$,
$\zeta_k = \sum \limits_{n=1}^{\infty} 1/n^k$ for the
Riemann $\zeta$-function and $a_4 = \sum \limits_{n=1}^{\infty} 1/(2^nn^4)
     \equiv {\rm Li}_4(1/2)$ in what follows.

\subsection{Master integral $I_{10}$}

 Let us consider the master integral $I_{10}$. 
 Our basic approach to computing 
 on-shell master integrals is to perform a large mass expansion 
 which yields an explicit power series representation for the on-shell 
 integrals.  The advantage of this approach is 
 that in this way most multi-loop 
 integrations can be performed rather easily, and the resulting 
 representation 
 of the on-shell master integral (expanded around the space-time 
 dimension $D=4$) takes the form of a nested harmonic sum which 
 can be reduced to known mathematical constants.

 This approach has been applied in earlier works \cite{largemassUntrunc}
 to obtain exact results for several on-shell integrals.  It relies
 heavily on an ability to  
 reduce specific classes of nested sums.
 A similar  problem of reducing harmonic sums appears in calculations for 
 deep-inelastic lepton-nucleon scattering, where one also deals with an 
  infinite expansion
 (the light-cone expansion instead of the large mass expansion that we employ
 here). Collections of summation identities of 
 the type that are needed for the 
 present work can be found in the literature 
 \cite{finitesums1,infinitesums1,infinitesums2,infinitesums3,sumsjos}. 
 
 For $I_{10}$ we start by taking the squared masses of the two particles inside
the loop $M^2$ to be different from the square of the external momenta
 $p^2=-m^2$.  Eventually, we are interested in the result 
 for $M^2=m^2$.

The expression for  Euclidean integral $I_{10}$ reads:
\begin{equation}
     \label{I10offshellfirst}
    I_{10} =
  \int\int\int \frac{{\rm d}^Dk_1\; {\rm d}^Dk_2\;{\rm d}^Dk_3\;}{
     (k_1+p)^2\; (k_2^2+M^2)\; (k_3^2+M^2)\; (k_1+k_2)^2 \;(k_3+k_2)^2}
 \end{equation} 
 and we want to perform a systematic expansion in $(p^2/M^2)$. 
  In  the present case this can be achieved by making an ordinary 
 Taylor expansion in the  external momentum $p$ in the integrand
\begin{eqnarray} \label{I10taylor}
 \frac{1}{[(k_1+p)^2]^{\alpha_1}} & \rightarrow  & \frac{1}{(k_1^2)^{\alpha_1}}
           \sum_{i=0}^\infty
           (-1)^i  \left( \frac{ 2k_1\cdot p +p^2}{k_1^2} \right)^i
           \frac{ \Gamma(\alpha_1+i)}{ \Gamma(\alpha_1)\; i!} .
\end{eqnarray}
As a side remark,  let us 
  note that for integral $I_{10}$ the same large mass expansion 
 can be obtained via a Mellin-Barnes representation for the two massive 
 lines \cite{mellinbarnes}. The Mellin-Barnes method 
  can be used successfully here because 
 the underlying massless integral with arbitrary complex
 powers of the massless lines can be evaluated in a closed form in terms 
 of a product of Euler $\Gamma$-functions. 
 However the diagrammatic large mass expansion as we use it here
 (i.e. the expansion via a sequence of Taylor expansions in the integrand)
  can be used also for integrals where the underlying
 massless integral is not known for arbitrary complex powers of 
 the propagators, and the diagrammatic approach to a large mass expansion 
 is therefore more suitable for certain complicated integrals,
 even though it seems technically more involved for  simple topologies.

  After performing the Taylor expansion in the integrand 
  in Eq. (\ref{I10offshellfirst})
  one is left with a 3-loop vacuum (bubble) integral with
  a tensor numerator involving powers of $(k_1\cdot p)$. This numerator
  is simplified after a bubble tensor reduction:
  \begin{equation}
 (k_1\cdot p)^{n}\rightarrow \left \{
     \begin{array}{cc}
         (k_1^2 p^2)^{\frac{n}{2}}
       \; \displaystyle \frac{n!}{2^{n}\left(n/2\right)!} \;
          \frac{\Gamma\left(D/2\right)}
               {\Gamma\left(D/2+n/2\right)},
       &{\rm for}~n~{\rm even},\\
         0, &{\rm for}~n~{\rm  odd},
     \end{array} \right.
 \end{equation}
  so that Eq.(\ref{I10taylor}) effectively becomes:
\begin{eqnarray}
 \frac{1}{[(k_1+p)^2]^{\alpha_1}} & \rightarrow  & \frac{1}{(k_1^2)^{\alpha_1}}
       \sum_{i=0}^\infty
       \left( \frac{p^2}{k_1^2} \right)^i
        \frac{\Gamma(\alpha_1+i)\; \Gamma(\alpha_1+i+1-D/2) \Gamma(D/2)}{ 
        \Gamma(\alpha_1)\; i!\; \Gamma(\alpha_1+1-D/2)\; \Gamma(i+D/2) }.
\end{eqnarray}
  The 3-loop vacuum bubble integral can now be evaluated 
  in terms of $\Gamma$-functions 
\ba
&&  \int\int\int \frac{{\rm d}^Dk_1\; {\rm d}^Dk_2\;{\rm d}^Dk_3\;}{
     (k_1^2)^{\alpha_1}(k_2^2+M^2)\;
        (k_3^2+M^2)\;[(k_1-k_2)^2]
        \;[(k_3-k_2)^2]} = 
\nonumber \\
&&      (-1)^{\alpha_1}\; \pi^{(3D/2)} M^{3D-8-2\alpha_1} \;
   \frac{ \Gamma(-3D/2+4+\alpha_1) \Gamma^3(D/2 - 1 ) \Gamma(2 - D/2)
         }{ \Gamma(D/2-1+\alpha_1) \Gamma(D/2) }
\nonumber \\
&&            \times \Biggl[
       \frac{ \Gamma^2(3 - D)}{ \Gamma(6 - 2D)}
    - 2 \sum_{j=1}^{\alpha_1}
      \frac{\Gamma(D/2-2+j) \Gamma(2+j-D) \Gamma(2 - D/2) 
           }{(j-1)!\;  \Gamma(-3D/2+4+j)  \Gamma( -1 + D/2) }
             \Biggr],    
\ea
  where $\alpha_1$ is any non-negative integer.
  In this way one obtains an explicit representation for $I_{10}$ 
   as an infinite power series in $(p^2/M^2)$.
  To obtain the result on the mass shell, where $m^2/M^2 = 1$,  
  we have to perform the infinite sum over the coefficients of 
  the power series.
  This task becomes much easier once we expand the obtained 
  representation for $I_{10}$ around $D=4$. For this purpose the 
$\Gamma$-functions are expanded as,
\begin{equation} \label{expandGamma}
        \Gamma(n+1+\varepsilon) = n!\; \Gamma(1+\varepsilon) \left[
              1+ \varepsilon S_1(n) + \frac{\varepsilon^2}{2}\left(
                   S_1^2(n)-S_2(n)\right) + {\cal O}(\varepsilon^3)\right],
\end{equation}
 where the harmonic sums $S_k$ are 
 defined as \begin{equation}
      S_k(n) = \sum_{i=1}^{n} \frac{1}{i^k}.
\end{equation}
 Expanding in  $\varepsilon$ we obtain the following off-shell 
 expression for $I_{10}$, written in a form that allows a straightforward 
  reduction of the remaining infinite sum as soon as we go on the mass shell
\ba
&& I_{10} = -\frac{C(\varepsilon)}{M^{-2+6\varepsilon}} \Biggl\{
 \frac{1}{ 3 \varepsilon^3}
       + \frac{2}{\varepsilon^2}
       + \frac{1}{\varepsilon} \left( \frac{25}{3} + 2\zeta_2 \right)
               + \left( 30 + 8\zeta_2 - \frac{22}{3}\zeta_3 \right)
       + \varepsilon  \biggl( \frac{301}{3} + 22\zeta_2
       - 24\zeta_3 + 33\zeta_4 \biggr)
\nonumber  \\
&&       + \varepsilon^2 \left( 322 - 4\zeta_2\zeta_3 + 52\zeta_2
                   - \frac{130}{3}\zeta_3 + 108\zeta_4 - 102\zeta_5 \right)
   +\frac{m^2}{M^{2}}
      \biggl[
           -\frac{1}{3 \varepsilon^2}
          - \frac{5}{6 \varepsilon}
         +  \left(  - \frac{19}{12} - \zeta_2 \right) 
\nonumber  \\
&&       + \varepsilon \left(  - \frac{23}{8} - 2\zeta_2 + \frac{7}{3}\zeta_3
                        \right)
       + \varepsilon^2 \left(  - \frac{85}{16} - \frac{7}{4}\zeta_2
               + \frac{10}{3}\zeta_3 - \frac{21}{2}\zeta_4 \right)
            \biggr]
          \Biggr\} 
\nonumber   \\
&&   -  \frac{C(\varepsilon)}{M^{-2+6\varepsilon}} \sum_{k=1}^{\infty}
       \left(\frac{m^2}{M^2}\right)^{k+1}  \Biggl\{
   \frac{1}{\varepsilon } \biggl[
         -\frac{1}{k}+\frac{1}{k+2} +\frac{2}{(k+1)^2}   \biggr]
     + \varepsilon^0\biggl[ 
        -\frac{10}{k} 
            -\frac{1}{1+k} +\frac{11}{2+k} 
          -5\frac{S_1(k)}{k}
\nonumber      \\
&&      -4 \frac{S_1(1+k)}{1+k}
          +9\frac{S_1(2+k)}{2+k}
           + \frac{3}{k^2}
        + \frac{12}{(1+k)^2}
           -\frac{10}{(1+k)^3} 
      +12 \frac{S_1(1+k)}{(1+k)^2}
      -\frac{8}{(2+k)^2}
\nonumber \\
&&       -\frac{2}{(2+k)^3}
      +2\frac{S_1(2+k)}{(2+k)^2}
          \biggr]
       + \varepsilon \biggl[
         \frac{-64-3\zeta_2}{k}
         + \frac{-13-4\zeta_2}{k+1}
           +\frac{77+7\zeta_2}{k+2}
       -13\frac{S^2_1(k)}{k}
\nonumber \\
&&       +33\frac{S^2_1(k+2)}{k+2}
             -20\frac{S^2_1(k+1)}{k+1} 
      - 50\frac{S_1(k)}{k}
          -38\frac{S_1(k+1)}{k+1} 
          +88 \frac{S_1(k+2)}{k+2}
         +5\frac{S_2(k)}{k} 
\nonumber \\
&&              +12 \frac{S_2(k+1)}{k+1} -17 \frac{S_2(k+2)}{k+2}
         +\frac{50+8\zeta_2}{(k+1)^2} 
         +\frac{-77+2\zeta_2}{(k+2)^2}
               +76\frac{S_1(k+1)}{(k+1)^2}
             + \frac{38}{(k+1)^4}
\nonumber  \\
&&             +15\frac{S_1(k)}{k^2}
          -\frac{9}{k^3}
          +\frac{30}{(k+2)^3}
                +\frac{12}{(k+2)^4}
           +10 \frac{S^2_1(k+2)}{(k+2)^2}
       +36\frac{S^2_1(k+1)}{(k+1)^2}
             -\frac{60}{(k+1)^3}
\nonumber  \\
&&              +\frac{30}{k^2}
        -56\frac{S_1(k+1)}{(k+1)^3}
            -16\frac{S_2(k+1)}{(k+1)^2}
            -45\frac{S_1(k+2)}{(k+2)^2}
           -16\frac{S_1(k+2)}{(k+2)^3} 
             -6 \frac{S_2(k+2)}{(k+2)^2}
             \biggr]
\nonumber  \\
&&   + \varepsilon^2 \biggl[
          \frac{7\zeta_3-30\zeta_2- 336}{k}
          +\frac{20\zeta_3-36\zeta_2-103}{k+1}
          +\frac{-27\zeta_3+66\zeta_2+439}{k+2}
      +\frac{(-320-16\zeta_2)S_1(k)}{k} 
\nonumber \\
&&              +\frac{(-236-32\zeta_2)S_1(k+1)}{k+1}
            +\frac{(556+48\zeta_2)S_1(k+2)}{k+2}
       -130\frac{S^2_1(k)}{k} -178\frac{S^2_1(k+1)}{k+1}
\nonumber \\
&&          +308\frac{S^2_1(k+2)}{k+2}
       -\frac{70}{3}\frac{S^3_1(k)}{k} -\frac{152}{3}\frac{S^3_1(k+1)}{k+1} 
         +74\frac{S^3_1(k+2)}{k+2}
       +50\frac{S_2(k)}{k} +104\frac{S_2(k+1)}{k+1}
\nonumber \\
&&        -154\frac{S_2(k+2)}{k+2}
      -\frac{26}{3}\frac{S_3(k)}{k}  -\frac{64}{3}\frac{S_3(k+1)}{k+1}
           +30\frac{S_3(k+2)}{k+2}
      -4\frac{S_{2,1}(k)}{k}  -16\frac{S_{2,1}(k+1)}{k+1}
\nonumber \\ 
&&             +20\frac{S_{2,1}(k+2)}{k+2}
      +28 \frac{S_1(k)S_2(k)}{k}
             +80\frac{S_1(k+1)S_2(k+1)}{k+1}
              -108\frac{S_1(k+2)S_2(k+2)}{k+2}
\nonumber  \\
&&         +\frac{9\zeta_2+192}{k^2}
     +\frac{-24\zeta_3+52\zeta_2+186}{(k+1)^2} 
              +\frac{-36\zeta_2-244}{(k+1)^3}
      +\frac{-10\zeta_3-29\zeta_2-479}{(k+2)^2}
      +\frac{-12\zeta_2+337}{(k+2)^3}
\nonumber    \\
&&       +39\frac{S^2_1(k)}{k^2} +236\frac{S^2_1(k+1)}{(k+1)^2}
       -129\frac{S^2_1(k+2)}{(k+2)^2} -56\frac{S^2_1(k+2)}{(k+2)^3}
         -160\frac{S^2_1(k+1)}{(k+1)^3}
        +150\frac{S_1(k)}{k^2}
\nonumber  \\
&&       -45\frac{S_1(k)}{k^3}
        -15\frac{S_2(k)}{k^2}
      -340\frac{S_1(k+1)}{(k+1)^3} +204\frac{S_1(k+1)}{(k+1)^4}
         -108 \frac{S_2(k+1)}{(k+1)^2}
      +68 \frac{S_2(k+1)}{(k+1)^3}
\nonumber  \\ 
&&         +28 \frac{S_3(k+1)}{(k+1)^2}
         +16 \frac{S_{2,1}(k+1)}{(k+1)^2}
       +157\frac{S_1(k+2)}{(k+2)^3}  +74\frac{S_1(k+2)}{(k+2)^4}
       +\frac{27}{k^4}  -\frac{90}{k^3} +\frac{228}{(k+1)^4}
\nonumber  \\
&&            -\frac{130}{(k+1)^5}
      +\frac{(330+48\zeta_2)S_1(k+1)}{(k+1)^2}
      +\frac{(-480+16\zeta_2)S_1(k+2)}{(k+2)^2}
      -40\frac{S_1(k+2)S_2(k+2)}{(k+2)^2}
\nonumber    \\
&&      -96\frac{S_1(k+1)S_2(k+1)}{(k+1)^2}
          -\frac{98}{(k+2)^4} 
          -\frac{50}{(k+2)^5}
      +55\frac{S_2(k+2)}{(k+2)^2} +28\frac{S_2(k+2)}{(k+2)^3} 
\nonumber   \\
&&       +\frac{32}{3}\frac{S_3(k+2)}{(k+2)^2}
              +8\frac{S_{2,1}(k+2)}{(k+2)^2}
        +72\frac{S^3_1(k+1)}{(k+1)^2}
           +\frac{76}{3} \frac{S^3_1(k+2)}{(k+2)^2}
         \biggr] \biggr\}  + O(\varepsilon^3).
     \label{I10fulloffshell}
  \ea
 The nested harmonic sums $S_{n,...,m}(i)$
 appearing in this expression are defined recursively via
      \begin{equation}
  S_{k,n,...,m}(i) = \sum_{j=1}^{i} \frac{ S_{n,...,m}(j)}{j^k}.
      \end{equation}
 Finally, in the on-shell limit $m^2=M^2=1$, the infinite sum over $k$ in 
  Eq.(\ref{I10fulloffshell}) is performed to yield the following result
\ba
    \label{I10fullonshell}
I_{10} && =  C(\ep) \left (
       - \frac {1}{3\ep^3}
       - \frac {5}{3\ep^2} 
       - \frac {1}{\ep} \left (  4 + \frac {2}{3}\pi^2 \right )
       + \frac {10}{3} - \frac {26}{3}\zeta_3 - \frac {7}{3}\pi^2
       + \ep \left ( \frac {302}{3} - \frac {94}{3}\zeta_3 
           - \pi^2 - \frac {35}{18}\pi^4 \right )
\right. \nonumber \\
&& \left.
       - \ep^2  \left (  -734 +\frac {76}{3} \pi^2\zeta_3 
       -\frac {101}{3}\pi^2 
      +20\zeta_3 +\frac {551}{90}\pi^4 +462\zeta_5 \right ) + {\cal O}(\ep^3)
\right ).
\ea
The infinite sums that one needs to go from 
  Eq.(\ref{I10fulloffshell}) to  Eq.(\ref{I10fullonshell}) are
given in Appendix B.

\subsection{Master integral $I_{5}$}

Let us now consider a calculation of the constant $C_1$ 
(see Ref.\cite{Laporta}).
The easiest way to extract it is to consider 
the integral $I_5$ of Ref.\cite{Laporta} (see Fig.2). We introduce
$\Pi_1=k_1^2, \Pi_3=(k_1-k_2)^2, \Pi_4 = k_1^2 + 2pk_1, 
\Pi_5 = k_2^2 + 2pk_2, \Pi_6=k_3^2+M^2,\Pi_7=(k_2+k_3)^2+M^2$
with $p^2=-m^2$ and consider
\be
I_{5}(m,M) = \int \frac {{\rm d}^Dk_1 {\rm d}^Dk_2  {\rm d}^Dk_3}
 {\Pi_1 \Pi_3 \Pi_4 \Pi_5  \Pi_6 \Pi_7}
\ee
for $m=M$. The result of Ref.\cite{Laporta} reads:
\ba
I_5(m,m)&&=C(\ep) \left \{
\frac {1}{6\ep^3}  + \frac {3}{2\ep^2} + \frac {1}{\ep} 
\left ( -\frac {\pi^2}{3} + \frac {55}{6} \right ) 
-\frac {4}{45} \pi^4 - \frac {14}{3}\zeta(3) -\frac {7}{3} \pi^2
+\frac {95}{2} 
\right.
\nonumber \\ 
&& \left.
+ \ep \left (
-\frac {2}{9} \pi^4 - 44\zeta_3 - \frac {29}{3} \pi^2
+\frac {1351}{6} + 2C_1 \right ) \right \},
\label{i5rem}
\ea
consequently, in order to extract $C_1$ we have to compute $I_5$ to 
order ${\cal O}(\ep)$. 

For this purpose, it turns out to be useful to derive a suitable 
integral representation for the vacuum polarization subdiagram in 
$I_5$. This representation can be obtained by subsequent application
of  dispersion  relation to ``vacuum polarization'' subgraph
and the Mellin-Barnes transformation. Let us first write a
dispersion representation for the vacuum polarization subgraph:
\be
\int \frac{ {\rm d}^Dk_3} {\Pi_6 \Pi_7} = 
\frac{1}{\pi} \int \limits_{4M^2}^{\infty} 
\frac {{\rm d}\lambda^2}{\lambda^2 + k_2^2} 
\frac {\pi^{7/2-\ep}}{2^{1-2\ep}\Gamma(3/2-\ep)}
\lambda^{-2\ep}
\left ( 1 - \frac {4M^2}{\lambda^2} \right )^{1/2-\ep}.
\label{e13}
\ee
Note that the above expression is written for an Euclidean 
momenta $k_2^2$.
We then use the Mellin-Barnes representation for the 
propagator $1/( k_2^2 + \lambda^2)$:
\be
\frac {1}{k_2^2 + \lambda^2} = \frac {1}{2\pi i} 
\int \limits_{C}^{}
{\rm d} \sigma {\Gamma(-\sigma)\Gamma(1+\sigma)}
 (k_2^2)^{\sigma} (\lambda^2)^{-1-\sigma},
\ee
where the integration is performed over the contour $C$ in 
the complex $\sigma$ plane which goes from  complex infinity
to complex infinity and passes the real axes between 
the poles of $\Gamma(-\sigma)$ 
and $\Gamma(1+\sigma)$. It turns out to be more convenient
to shift the integration contour to the right,  
crossing the singularity at $\sigma=0$, so that the new integration 
contour $\tilde C$ crosses the  real axes at e.g. $\sigma=1/2$. We obtain:
\be
\frac {1}{k_2^2 + \lambda^2} = \frac {1}{\lambda^2} + 
\frac {1}{2\pi i} 
\int \limits_{\tilde C}^{}
{\rm d} \sigma {\Gamma(-\sigma)\Gamma(1+\sigma)}
 (k_2^2)^{\sigma} (\lambda^2)^{-1-\sigma}.
\label{repr}
\ee
Let us study two terms in that equation separately. The first term 
produces a factorized expression where the integral over 
$\lambda$ can be carried out independently of the integration 
over $k_1$ and $k_2$. The remaining two loop integral is 
an on-shell integral and for this reason can be easily done.
The result of this calculation reads
(we take the limit $m=M$):
\ba
I_{5a}(m,m)&&=C(\ep) \left \{
\frac {1}{6\ep^3}  + \frac {3}{2\ep^2} + \frac {1}{\ep} 
\left ( -\frac {\pi^2}{3} + \frac {55}{6} \right ) 
-\frac {8}{3} \zeta_3 - \pi^2
+\frac {43}{2} 
\right.
\nonumber \\ 
&& \left.
+ \ep \left (
-\frac {19}{45} \pi^4 - 16\zeta_3 - \frac {5}{3} \pi^2
+\frac {775}{6} \right ) \right \}.
\ea

In order to compute the contribution of the second term in Eq.(\ref{repr})
we insert it into  dispersion integral Eq.(\ref{e13}) 
and integrate over $\lambda$.
We obtain:
\ba
I_{5b}(m,M) =&& \frac {1}{2\pi i}
 \frac {\pi^{5/2-\ep}}{2^{1-2\ep}\Gamma(3/2-\ep)}
 \int \limits_{\tilde C}^{}
{\rm d} \sigma \Gamma(-\sigma) 
 \Gamma(1+\sigma) {\rm B}\left (\sigma+\ep, \frac {3}{2}-\ep \right )
 (4M^2)^{-\ep-\sigma} 
\nonumber \\
&& \times \int \frac {{\rm d}^Dk_1 {\rm d}^Dk_2  }
 {\Pi_1 \Pi_3 \Pi_4 \Pi_5} (\Pi_2)^{\sigma}, 
\label{i5b}
\ea
where $\Pi_2 = k_2^2$.

To proceed further, we need to perform the integrals over $k_1$ and
$k_2$. The easiest way to do that is to utilize the integration-by-parts
identities. It turns out that there is an identity that allows to remove 
either $\Pi_3$ or $\Pi_4$ or $\Pi_5$ in which case the integrals become
computable in a closed form for arbitrary $\sigma$. To write down this
identity, we introduce 
\be
I(a_1,a_2,a_3,a_4,a_5) = 
\int \frac {{\rm d}^Dk_1 {\rm d}^Dk_2  }
 {\Pi_1^{a_1} \Pi_2^{a_2} \Pi_3^{a_3} \Pi_4^{a_4} \Pi_5^{a_5}}. 
\ee
The general  form of the identity is\cite{ac}:
\ba
I(\{a_i\}) &&= 
  \frac {-1}{2 (2D-2\sum_{i=1}^{5}a_i +a_4+a_5)}
  \Big \{
  \frac {2 (D-1-a_3 - a_1)(D - 1 - a_3 - a_2)}{(D-1-2a_3)}{\bf 3}^-
\nonumber \\
&&  
  -\frac {(D-1-a_3-a_2)}{(D-1-2a_3)}{\bf 4}^+{\bf 5}^- {\bf 1}^-
  -\frac {(D-1-a_3-a_1)}{(D-1-2a_3)} {\bf 5}^+{\bf 4}^- {\bf 2}^-
\nonumber \\
&& 
  + a_3 {\bf 3}^+({\bf 4}^- -  {\bf 5}^-   )({\bf 2}^- - {\bf 1}^-   )
  -(a_2-1) {\bf 5}^-  - (a_1-1) {\bf 4}^-
  \Big \} I(\{a_i\}),
\ea
where as usual ${\bf 1^\pm}~I(a_1,...) = I(a_1 \pm 1,...)$ and similar
for other operators.

It is now easy to see that if we apply that recurrence relation 
to the integral $I(1,-\sigma,1,1,1)$, we immediately obtain 
a set of integrals that can be easily computed
since the recurrence relation removes either one of the massive
lines (in which case there appears a massless two point subgraph)
or the line $\Pi_3$ (in which case the integral becomes a product of
two one loop integrals). Therefore, we end up with the expression 
for $I(1,-\sigma,1,1,1)$ written in terms of $\Gamma$-functions:
\ba
&&I(1,-\sigma,1,1,1) =  \frac {1}{2 +2\sigma -4\ep}
       \Big [ -\frac {1+\sigma}{2}
      S_2\left (1,-\sigma,1,1 \right )
           - \frac{1}{2} S_2(1,2, - 1 - \sigma,1) 
\nonumber \\
&&  
       + \frac {1}{2} S_2(2, - 1 - \sigma,1,1)
 -\frac {1}{2} S_2(2,-\sigma,0,1)  + \frac {1}{1-2\ep}  \Big ( 
       \left (1 - \ep + \frac {1}{2}\sigma \right )
        S_2(1,-\sigma,0,2) 
\nonumber \\
&&  
 + \left ( \frac {1}{2} - \ep \right ) S_2(1,1, - 1 -\sigma,2) 
        + \left ( - 2 + 2\ep\sigma + 6\ep - 4\ep^2 
        -\sigma \right )S_1(1,1)~S_1(-\sigma,1)
 \Big )
\Big ],
\ea
where the functions $S_2$ and $S_1$ are defined as follows:
\ba
&&S_2(a_1,a_2,a_3,a_4)=O_1(a_1,a_2)~S_1(a_1+a_2+a_3-2+\ep,a_4),
\nonumber \\
&&
O_1(a_1,a_2)=\frac {\Gamma(2-\ep-a_1)\Gamma(2-\ep-a_2)\Gamma(a_1+a_2-2+\ep)}
{\Gamma(a_1) \Gamma(a_2) \Gamma(4-2\ep-a_1-a_2)},
\nonumber \\
&&S_1(a_1,a_2)=\frac {\Gamma(a_1+a_2-2+\ep)\Gamma(4-2\ep-2a_1-a_2)}
{\Gamma(a_2) \Gamma(4-2\ep-a_1-a_2)}.
\ea

We use these expressions in Eq.(\ref{i5b}) and integrate over 
$\sigma$ by taking the residues located to the right of the 
integration contour, i.e. 
at the points $\sigma = n,~n+\ep,~n+2\ep$,
where $n \ge 1$ is an integer\footnote{It is interesting to note that 
different poles in $\sigma$ correspond to contributions of different expansion 
regimes if $I_5$ is expanded  in $m/M$ according to the rules
of the large mass expansion (for a review see e.g. \cite{smirnov}).}. 
Finally we arrive at the 
representation for $I_5$ written as an infinite series:
\be
I_{5b}(m,M) = \sum \limits_{n}^{\infty} d_n 
 \left ( \frac {m^2}{M^2} \right )^n.
\ee
The expression for the coefficient $d_n$ is too long to be presented
here; the essential point, however, is that $d_n$
can be expressed through harmonic 
sums and the summation for $m=M$  can be  carried out
in a way similar to the one described in the previous subsection.
Our final  result for $I_{5b}(m,m)$ reads:
\ba
I_{5b}(m,m)&& = -\frac {4}{45}\pi^4 - 2 \zeta(3) - \frac {4}{3}\pi^2
+26 
\nonumber \\
&&
+ \ep \left (
 96 + \frac {25}{6}\pi^2\zeta_3 + 4\pi^2\log 2 - \frac {26}{3}\pi^2 
- 26\zeta_3  - \frac {31}{90}\pi^4 - \frac {49}{2} \zeta_5
\right ). 
\ea

Combining the results for $I_{5a}$ and $I_{5b}$ and comparing with 
Eq.(\ref{i5rem}), we obtain the constant $C_1$:
\be
C_1 =  \zeta_3
-\frac {1}{3}\pi^2-\frac {49}{180}\pi^4+\frac {25}{12}\pi^2\zeta_3
 +2\pi^2 \log 2 -\frac {49}{4}\zeta_5.
\ee

\subsection{Master integral $I_{18}$}

For integral $I_{18}$ we proceed as for $I_{10}$ and perform an
 expansion in $p^2/M^2$ of a more general Euclidean integral
\be
     \label{I18offshellfirst}
    I_{18} =
  \int\int\int \frac{{\rm d}^Dk_1\; {\rm d}^Dk_2\;{\rm d}^Dk_3\;}{
     (k_1+p)^2\; (k_3+p)^2\; (k_1-k_2)^2\; (k_3-k_2)^2\; (k_1^2+M^2)\;
      (k_2^2+M^2)\; (k_3^2+M^2)}.
 \ee
  As is the case for $I_{10}$, the expansion of $I_{18}$
 can be performed\footnote{We should 
  note that a large mass expansion of integrals
  $I_{18}$ and $I_{10}$ is rather special since no contributions arise 
  other than the Taylor expansion of the whole integrand in $p$.
  However, for other types of integrals without a through-going 
  massive line the combinatorics of the large mass expansion gives
  rise to several more subgraphs that must be Taylor expanded.}
  by making an ordinary Taylor expansion in the external momentum 
  $p$ in the integrand of Eq. (\ref{I18offshellfirst}),
  i.e. by expanding both $1/(k_1+p)^2$ and $1/(k_3+p)^2$ in small $p$ as 
 in Eq.(\ref{I10taylor}).

 The resulting integral is formally a 3-loop vacuum bubble integral that, 
after the necessary tensor reductions and partial fractioning
 of massive and massless lines, can be evaluated in terms
 of $\Gamma$-functions using Eqs.(\ref{1loopbubble}-\ref{bubblem00})  
  in the Appendix A. 
 In this way one obtains an explicit representation for $I_{18}$ as an
 infinite series in $p^2/M^2$ with coefficients being a
 nested sum over a product of $\Gamma$-functions. After expanding the
 expressions in $\varepsilon$ as in Eq.(\ref{expandGamma})
 and writing the nested sums in a form that allows a straightforward
 reduction on the mass shell we obtain a fairly simple expression 
\ba
  I_{18} &&= 
     \frac{C(\varepsilon)}{M^{2+6\varepsilon}} \sum_{k=1}^{\infty}
       \left(\frac{m^2}{M^2}\right)^{k}  \Biggl[
           2\frac{\zeta_3}{k^2}   -2\frac{\zeta_2}{k^3} 
+4\zeta_2\frac{S_1(k)}{k^2}
        + 2\frac{S_1(k)}{k^4} -2\frac{S^2_1(k)}{k^3}  
\nonumber  \\
        && +4\frac{S_1(k)S_2(k)}{k^2}  -4\frac{S_{2,1}(k)}{k^2}
        + O(\varepsilon) \Biggr].
\ea
 The on-shell result is now easily recovered by putting
  $m^2=M^2=1$ and using the expressions for the infinite sums
    Eqs. (\ref{sumlistbegin}-\ref{sumlistend}). We finally obtain
   \be
   \label{I18fullonshell}
  I_{18} = 
   C(\varepsilon)\Biggl[
            2 \pi^2 \zeta_3 - 5\zeta_5 + O(\varepsilon) \Biggr].
  \ee

\vspace*{1cm}
Having derived  the last  unknown contribution for the three loop
on-shell two-point master integrals, we are able to write down a complete
set of master integrals for the three loop on-shell two point functions
without unknown constants. The list  updates the 
result of Ref. \cite{Laporta} and can be found in Appendix C.

\section{Three loop QCD renormalization constants in the on-shell scheme}

Performing an explicit computation along the lines outlined in the previous 
Sections, we are able to derive an expression for the QCD on-shell 
renormalization constants up to the three loop order.  
We work here in a general covariant gauge with a gauge fixing 
parameter $\xi$; the gluon propagator (omitting the color indices) 
therefore reads:
\be  \label{definexi}
D_{\mu \nu}(p) =  -\frac {i}{p^2} 
\left ( g_{\mu \nu} - \xi \frac {p_\mu p_\nu}{p^2}
  \right ).
\ee

In our previous article \cite{uns} we have given a relation 
between the pole and the $\overline {\rm MS}$ mass, from where 
the value of $Z_m$ can be determined. Here we give the 
result for that renormalization constant explicitly.
We write:
\be
Z_m = 1 + a_0 C_F \left (  -\frac {3}{4\ep}-\frac {1}{1-2\ep}  \right ) 
 + a_0^2 C_F Z_m^{(2)} + a_0^3 C_F Z_m^{(3)},
\ee
where 
$$
a_0 = \frac {\alpha_s^{(0)}}{\pi}
\frac {\Gamma(1+\ep) m^{-2\ep}}{(4 \pi)^{-\ep}},
$$
and $\alpha_s^{(0)}$ is the bare coupling constant and $Z_m^{(2)}$ and 
$Z_m^{(3)}$ are the two and the three loop coefficients respectively:
\ba
Z_m^{(2)} && = 
C_F\, d^{(2)}_1  + C_A\, d^{(2)}_2 
+ T_RN_L\,d^{(2)}_3
+T_RN_H \, d^{(2)}_4,
\nonumber \\
Z_m^{(3)} && = 
C_F^2\, d^{(3)}_1
 + C_FC_A \, d^{(3)}_2 
 + C_A^2\, d^{(3)}_3
 + C_F T_R N_L \, d^{(3)}_4
 + C_F T_R N_H \, d^{(3)}_5
 + C_A T_R N_L \, d^{(3)}_6
\nonumber \\
&& + C_A T_R N_H \, d^{(3)}_7
 + T_R^2 N_L N_H \, d^{(3)}_8
 + T_R^2 N_H^2 \, d^{(3)}_{9}
 + T_R^2 N_L^2 \, d^{(3)}_{10}.
\ea
Here $C_F$ and $C_A$ are the Casimir operators of the fundamental and the
adjoint representation of the color gauge group (the group is SU(3)
 for QCD) and $T_R$ is the trace normalization of the
  fundamental representation.
  $N_L$ is the number of massless quark flavors and $N_H$ is the number of
  quark flavors with a pole mass equal to $m$.

Our results for the coefficients $d^{(i)}_{j}$ are:
\ba
d^{(2)}_1 =&& 
\frac {9}{32\ep^2} + \frac {45}{64\ep}  +  \frac {199}{128} 
 - \frac {3}{4}~\zeta_3 + \frac {1}{2}~\pi^2 \log 2 
 - \frac {5}{16} \pi^2 
\nonumber \\
&&+\ep \left ( \frac {677}{256} - \frac {33}{4}\zeta_3 + 3\pi^2\log 2 
- \pi^2 \log^2 2
          - \frac {55}{32}\pi^2 + \frac {7}{40}\pi^4 - 
 \frac {1}{2}\log^4 2 - 12 a_4 \right ),
\nonumber \\
d^{(2)}_2 =&&  
  - \frac {11}{32\ep^2}    - \frac {91}{64\ep}  - \frac {605}{128} 
 + \frac {3}{8}~\zeta_3 - \frac {1}{4}~\pi^2 \log 2 
 + \frac {1}{12}~\pi^2
\nonumber \\
&&+\ep \left (  - \frac {3799}{256} + \frac {13}{4}\zeta_3 - 
 \frac {3}{2}\pi^2 \log 2 + \frac {1}{2}\pi^2\log^2 2 + \frac {19}{48}\pi^2 
 - \frac {7}{80}\pi^4 + \frac {1}{4}\log^4 2 + 6 a_4 \right ),
\nonumber \\
d^{(2)}_3 =&& 
\frac {1}{8\ep^2}  + \frac {7}{16\ep} +  \frac {45}{32} + 
 \frac {1}{12}~\pi^2
+ \ep  \left (  \frac {279}{64} + \zeta_3 + \frac {7}{24}\pi^2  \right ),
\nonumber \\
d^{(2)}_4 =&& 
\frac {1}{8\ep^2} + \frac {7}{16\ep} +  \frac {69}{32} 
 - \frac {1}{6}~\pi^2 +
\ep \left ( \frac {463}{64} - \frac {7}{2}\zeta_3 + 
 \pi^2 \log 2 - \frac {5}{6}\pi^2 \right ),
\nonumber \\
d^{(3)}_1 =&&  - \frac {9}{128\ep^3}     - \frac {63}{256\ep^2}    
 + \frac {1}{\ep} \left ( - \frac {457}{512} 
 + \frac {9}{16}\zeta_3 - \frac {3}{8}~\pi^2 \log 2 
 + \frac {15}{64}~\pi^2 \right )
-\frac {14225}{3072}
\nonumber \\
&&
 - \frac {1}{16}\zeta_3\pi^2 
 + \frac {9}{8}\zeta_3 + \frac {5}{8}\zeta_5
+ 5\pi^2\log 2
+ \frac {5}{4}\pi^2\log^2 2 
          - \frac {731}{384}\pi^2 - \frac {73}{480}\pi^4 
           - \frac {1}{8}\log^4 2 - 3a_4,
\nonumber \\
d^{(3)}_2 =&& 
\frac {33}{128\ep^3}  +  \frac {1039}{768\ep^2}  + 
 \frac {1}{\ep} \left ( 
 \frac {2563}{512} - \frac {53}{32}\zeta_3 + \frac {53}{48}\pi^2\log 2
          - \frac {61}{96}\pi^2 \right )
\nonumber \\
&& 
+ \frac {144929}{9216} - \frac {19}{16}\zeta_3\pi^2 - \frac {2459}{96}\zeta_3 
 + \frac {45}{16}\zeta_5 
+ \frac {223}{36}\pi^2 \log 2 
    - \frac {221}{72}\pi^2 \log^2 2
\nonumber \\
&&
 - \frac {1477}{576}\pi^2 + \frac {4639}{8640}\pi^4 
 - \frac {167}{144}\log^4 2 - \frac {167}{6} a_4,
\nonumber \\
d^{(3)}_3 =&& 
- \frac {121}{576\ep^3}   - \frac {167}{108\ep^2}  + \frac {1}{\ep} 
 \left (  
 - \frac {81797}{10368} + \frac {11}{16}\zeta_3 
 - \frac {11}{24}~\pi^2 \log 2 + \frac {11}{72}~\pi^2  \right )
\nonumber \\
&&
- \frac {2188703}{62208} + \frac {51}{64}\zeta_3 \pi^2 
 + \frac {3059}{288}\zeta_3 
- \frac {65}{32}\zeta_5 
 - \frac {313}{72}\pi^2 \log 2
\nonumber \\
&&
 + \frac {11}{9}\pi^2\log^2 2 + \frac {553}{3456}\pi^2 - 
         \frac {3667}{17280}\pi^4 + \frac {11}{18}\log^4 2 
  + \frac {44}{3}a_4, 
\nonumber \\
d^{(3)}_4 =&& 
 - \frac {3}{32\ep^3}    - \frac {77}{192\ep^2}  
 +\frac {1}{\ep} \left (  - \frac {415}{384} + \frac {1}{4}\zeta_3 
 - \frac {1}{3}\pi^2 \log 2
          + \frac {7}{48}~\pi^2  \right )
  - \frac {2911}{2304} + \frac {169}{24}\zeta_3 
\nonumber \\
&&  - \frac {29}{9}\pi^2 \log 2
 + \frac {8}{9}\pi^2 \log^2 2
 + \frac {475}{288}\pi^2 - \frac {371}{2160}\pi^4 + \frac {4}{9}\log^4 2 
 + \frac {32}{3} a_4,
\nonumber \\
d^{(3)}_5 =&& 
 - \frac {3}{32\ep^3}    - \frac {77}{192\ep^2}  + \frac {1}{\ep} 
 \left (  - \frac {631}{384}
 + \frac {1}{4}\zeta_3 - \frac {1}{3}~\pi^2 
 \log 2 + \frac {1}{3}~\pi^2 \right )
 - \frac {8743}{2304} + \frac {71}{12}\zeta_3 
\nonumber \\
&& - \frac {67}{36}\pi^2\log 2
          + \frac {5}{9}\pi^2 \log^2 2
+ \frac {425}{432}\pi^2 
  - \frac {161}{2160}\pi^4 + \frac {4}{9}\log^4 2 + \frac {32}{3} a_4,
\nonumber \\
d^{(3)}_6 =&& 
\frac {11}{72\ep^3}  + \frac {877}{864\ep^2}  + \frac {1}{\ep} \left (
 \frac {3125}{648} + \frac {1}{6} \pi^2 \log 2 + \frac {7}{72} \pi^2  \right )
+ \frac {317381}{15552} + \frac {41}{144}\zeta_3 
\nonumber \\
&& + \frac {29}{18}\pi^2\log 2
          - \frac {4}{9}\pi^2 \log^2 2
+ \frac {73}{108}\pi^2 + \frac {29}{432}\pi^4 - \frac {2}{9}\log^4 2 
 - \frac {16}{3} a_4,
\nonumber \\
d^{(3)}_7 =&& 
\frac {11}{72\ep^3}  + \frac {877}{864\ep^2} + \frac {1}{\ep} 
\left ( \frac {502}{81} 
 + \frac {1}{6}\pi^2 \log 2 - \frac {13}{36}\pi^2  \right )
+\frac {473549}{15552} + \frac {1}{8}\zeta_3 \pi^2 
 - \frac {1345}{144}\zeta_3
\nonumber \\
&&          - \frac {5}{8}\zeta_5 + \frac {115}{18}\pi^2 \log 2 
 - \frac {5}{18}\pi^2\log^2 2 - \frac {707}{144}\pi^2 
   + \frac {1}{54}\pi^4 - \frac {2}{9}\log^4 2 - \frac {16}{3}a_4,
\nonumber \\
d^{(3)}_8 =&& 
 - \frac {1}{18\ep^3}   - \frac {17}{54\ep^2}  + \frac {1}{\ep}   
 \left (  - \frac {152}{81} + \frac {1}{18}~\pi^2 \right )
 - \frac {2032}{243} + \frac {17}{9}\zeta_3 
 - \frac {2}{3}\pi^2 \log 2 + 
         \frac {13}{27}\pi^2,
\nonumber \\
d^{(3)}_9 =&& 
- \frac {1}{36\ep^3}    - \frac {17}{108\ep^2}  
+ \frac {1}{\ep} \left ( - \frac {385}{324} + \frac {1}{9}~\pi^2 \right )
- \frac {5441}{972} + \frac {53}{18}\zeta_3 - \frac {2}{3}\pi^2\log 2 + 
         \frac {79}{135}\pi^2, 
\nonumber \\
d^{(3)}_{10} =&& 
 - \frac {1}{36\ep^3}   - \frac {17}{108\ep^2} + \frac {1}{\ep}   
 \left (  - \frac {223}{324} - \frac {1}{18}~\pi^2  \right )
  - \frac {2687}{972} - \frac {19}{18}\zeta_3 
 - \frac {17}{54}\pi^2. 
\ea
We have used here $\zeta_k = \sum \limits_{n=1}^{\infty} 1/n^k$ for the
Riemann $\zeta$-function and also 
$a_4 = \sum \limits_{n=1}^{\infty} 1/(2^nn^4)$.

It is more cumbersome to present the result for the wave function 
renormalization $Z_2$ since, as 
we mentioned already, it becomes gauge dependent at ${\cal O}(\alpha_s^3)$.
We parameterize  $Z_2$ as:
\be
Z_2 = 
1 + a_0 C_F \left (-\frac {3}{4\ep}-\frac {1}{1-2\ep}  \right ) 
 + a_0^2 C_F Z_2^{(2)} + a_0^3 C_F Z_2^{(3)},
\ee
where
\ba
Z_2^{(2)} && = 
C_F\, f^{(2)}_1  + C_A\, f^{(2)}_2 
+ T_RN_L\,f^{(2)}_3
+T_RN_H \, f^{(2)}_4,
\nonumber \\
Z_2^{(3)} && = 
C_F^2\, f^{(3)}_1
 + C_FC_A \, f^{(3)}_2 
 + C_A^2\, f^{(3)}_3
 + C_F T_R N_L \, f^{(3)}_4
 + C_F T_R N_H \, f^{(3)}_5
 + C_A T_R N_L \, f^{(3)}_6
\nonumber \\
&& + C_A T_R N_H \, f^{(3)}_7
 + T_R^2 N_L N_H \, f^{(3)}_8
 + T_R^2 N_H^2 \, f^{(3)}_{9}
 + T_R^2 N_L^2 \, f^{(3)}_{10}.
\ea

Our result for the coefficients $f^{(i)}_j$ reads:
\ba
f^{(2)}_1 && = \frac {9}{32\ep^2} + \frac {51}{64\ep}
 + \frac {433}{128} - \frac {3}{2}\zeta_3 + \pi^2\log 2 - \frac {13}{16}\pi^2 
\nonumber \\
&& +\ep \left ( \frac {211}{256} - \frac {147}{8}\zeta_3 
 + \frac {23}{4}\pi^2\log 2 - 2\pi^2 \log^2 2 - \frac {89}{32}\pi^2 
 + \frac {7}{20}\pi^4 - \log^4 2 - 24 a_4 \right ),
\nonumber \\
f^{(2)}_2 &&= - \frac {11}{32\ep^2} - \frac {101}{64\ep} 
 - \frac {803}{128} + \frac {3}{4}\zeta_3 - \frac {1}{2}\pi^2\log 2  
 + \frac {5}{16}\pi^2
\nonumber \\
&&+ \ep \left ( - \frac {4241}{256} + \frac {129}{16}\zeta_3 
 - \frac {23}{8}\pi^2\log 2 + 
         \pi^2\log^2 2 + \frac {41}{48}\pi^2 - \frac {7}{40}\pi^4 
 + \frac {1}{2}\log^4 2 + 12a_4 \right ),
\nonumber \\
f^{(2)}_3 &&= \frac {1}{8\ep^2} + \frac {9}{16\ep} 
+ \frac {59}{32} + \frac {1}{12}\pi^2
+ \ep \left ( \frac {369}{64} + \zeta_3 + \frac {3}{8}\pi^2 \right ),
\nonumber \\
f^{(2)}_4 &&=\frac {1}{4\ep^2} + \frac {19}{48\ep} 
+ \frac {1139}{288} - \frac {1}{3}\pi^2
  + \ep \left ( \frac {20275}{1728} - 7\zeta_3 
 + 2\pi^2\log 2 - \frac {19}{12}\pi^2 \right ),
\nonumber \\
f^{(3)}_1 && =  - \frac {9}{128\ep^3}     - \frac {81}{256\ep^2}  
 + \frac {1}{\ep} \left (  
- \frac {1039}{512} 
 + \frac {9}{8}\zeta_3 - \frac {3}{4}~\pi^2 \log 2 
 + \frac {39}{64}~\pi^2 \right )
 - \frac {10823}{3072} + \frac {1}{8}\zeta_3 \pi^2 
\nonumber \\
&& - \frac {187}{32}\zeta_3 
  - \frac {5}{16}\zeta_5 
+ \frac {685}{48}\pi^2\log 2 + 3\pi^2 \log^2 2
  - \frac {7199}{1152}\pi^2 - \frac {41}{120}\pi^4 - \frac {5}{12}\log^4 2 
 - 10a_4,
\nonumber \\
f^{(3)}_2 && =
\frac {33}{128\ep^3}   +  \frac {1217}{768\ep^2}  + \frac {1}{\ep} 
 \left (
 \frac {14887}{1536} - \frac {27}{8}\zeta_3 + \frac {53}{24}
 \pi^2\log 2
          - \frac {331}{192}\pi^2 \right )
\nonumber \\
&& +\frac {150871}{9216} - \frac {45}{16}\zeta_3 \pi^2 
 - \frac {9941}{192}\zeta_3 
 + \frac {145}{16}\zeta_5 
+ \frac {2281}{288}\pi^2 \log 2 
\nonumber \\
&& - \frac {499}{72}\pi^2 \log^2 2
 - \frac {1169}{576}\pi^2 + 
         \frac {20053}{17280}\pi^4 - \frac {319}{144} \log^4 2 - 
  \frac {319}{6} a_4,
\nonumber \\
f^{(3)}_3 && =
 - \frac {121}{576\ep^3}   - \frac {1501}{864\ep^2}  + \frac {1}{\ep} 
 \left (  
 - \frac {55945}{5184} + \frac {173}{128}\zeta_3 
 - \frac {11}{12}~\pi^2 \log 2 + \frac {55}{96}~\pi^2
 - \frac {\pi^4}{1080}
\right.
\nonumber \\
&& \left. +\xi \left ( -\frac {3}{256}\zeta_3 
                     +\frac {\pi^4}{4320} - \frac {1}{768}  
                 \right )  \right )
  - \frac {2551697}{62208} + \frac {127}{72}\zeta_3\pi^2 
        + \frac {14371}{576}\zeta_3 - \frac {37}{6}\zeta_5 
\nonumber \\
&&        - \frac {271}{36}\pi^2 \log 2
+ \frac {391}{144}\pi^2 \log^2 2 + \frac {1073}{3456}\pi^2
          - \frac {10811}{23040}\pi^4 + \frac {349}{288}\log^4 2 
 + \frac {349}{12}a_4
\nonumber \\
&& + \xi \left (
  - \frac {13}{768} + \frac {1}{144}\zeta_3\pi^2 
   - \frac {13}{256}\zeta_3 
          + \frac {7}{384}\zeta_5 - \frac {1}{256}\pi^2 
       + \frac {17}{27648}\pi^4 \right),
\nonumber \\
f^{(3)}_4 && =
 - \frac {3}{32\ep^3}   -\frac {103}{192\ep^2}  
 +\frac {1}{\ep} \left (  - \frac {351}{128} + \frac {3}{4}\zeta_3 
 - \frac {2}{3}\pi^2 \log 2
          + \frac {23}{48}\pi^2 \right )
 - \frac {3773}{2304} + \frac {413}{24}\zeta_3 
\nonumber \\
&& - \frac {58}{9}\pi^2\log 2
          + \frac {16}{9}\pi^2 \log^2 2 
+ \frac {905}{288}\pi^2
 - \frac {733}{2160}\pi^4 + \frac {8}{9}\log^4 2 
  + \frac {64}{3} a_4,
\nonumber \\
f^{(3)}_5 && =
- \frac {3}{16\ep^3}    - \frac {91}{192\ep^2} + \frac {1}{\ep} 
 \left (  - \frac {1525}{384}
 + \zeta_3 - \frac {2}{3}~\pi^2 \log 2 + \frac {19}{24}~\pi^2 \right )
 - \frac {78967}{6912} + \frac {5273}{288}\zeta_3 
\nonumber \\
&& - \frac {31}{9}\pi^2 \log 2 + \frac {5}{6}\pi^2 \log^2 2 
+ \frac {865}{648}\pi^2 
 - \frac {137}{720}\pi^4 + \frac {7}{6}\log^4 2 + 28a_4, 
\nonumber \\
f^{(3)}_6 && =
 \frac {11}{72\ep^3}  +  \frac {1069}{864\ep^2}  + \frac {1}{\ep} 
 \left ( 
 \frac {550}{81} -\frac {1}{4}\zeta_3  
 + \frac {1}{3} \pi^2 \log 2 - \frac {1}{18} \pi^2 \right )
+ \frac {416405}{15552} - \frac {145}{36}\zeta_3 + \frac {29}{9}\pi^2\log 2
\nonumber \\
&&  - \frac {8}{9}\pi^2\log^2 2 + \frac {7}{18} \pi^2 
 + \frac {29}{216}\pi^4 - \frac {4}{9}\log^4 2 - \frac {32}{3}a_4, 
\nonumber \\
f^{(3)}_7 && =
\frac {1}{\ep^3} \left ( \frac {15}{64} - \xi \frac {1}{192} \right )  
  + \frac {1}{\ep^2} \left ( 
 \frac {353}{288}+\xi \frac {1}{64} \right )  + \frac {1}{\ep} 
 \left ( \frac {503}{48}  
 + \frac {1}{3}\pi^2~\log 2 - \frac {59}{72}\pi^2 -\frac {1}{2}\zeta_3
              -\xi \frac {35}{576} \right )
\nonumber \\
&& +\frac {32257}{648} + \frac {11}{48}\zeta_3\pi^2 
- \frac {13301}{576}\zeta_3
          - \frac {15}{16}\zeta_5 + \frac {521}{36}\pi^2\log 2 
   - \frac {1}{3}\pi^2 \log^2 2
- \frac {13567}{1296}\pi^2 
\nonumber \\
&& + \frac {5}{72}\pi^4 - \frac {2}{3}\log^4 2 - 16a_4 
+ \xi \left (
 \frac {407}{1728} - \frac {1}{24}\zeta_3 \right ),
\nonumber \\
f^{(3)}_8 && =
  - \frac {1}{12\ep^3}    - \frac {7}{18\ep^2}    
 + \frac {1}{\ep} \left (  - \frac {31}{9} + \frac {1}{6}\pi^2  \right ) 
 - \frac {1168}{81} + 4\zeta_3 - \frac {4}{3}\pi^2\log 2 
 + \frac {7}{6}\pi^2,
\nonumber \\
f^{(3)}_9 && =
 - \frac {1}{12\ep^3}     - \frac {5}{36\ep^2}  + \frac {1}{\ep}     
 \left ( - \frac {131}{54} + \frac {2}{9}\pi^2 \right )
  - \frac {6887}{648} + 7\zeta_3 
 - \frac {4}{3}\pi^2 \log 2 
        + \frac {11}{10} \pi^2,
\nonumber \\
f^{(3)}_{10} && =
 - \frac {1}{36\ep^3}     - \frac {23}{108\ep^2}   
 + \frac {1}{\ep} \left (  - \frac {325}{324} - \frac {1}{18}\pi^2 \right )
 - \frac {4025}{972} - \frac {19}{18}\zeta_3 - \frac {23}{54}\pi^2.
\ea

Here the gauge parameter $\xi$ is defined in Eq.(\ref{definexi}) and
we again stress that the three loop contribution to $Z_2$ is gauge dependent. 
This fact is not too surprising by itself, since 
the wave function renormalization constants are known 
to be gauge dependent in general (as one can see for instance in 
the $\overline{\rm MS}$ scheme). The on-shell renormalization  
constant computed in dimensional regularization, however, unexpectedly
turned out to be gauge independent in two first orders in perturbation 
theory. This fact,  observed for the first time in \cite{broadhurst},
resulted in ``all orders gauge independence'' conjecture
by the authors of that reference. Our calculation shows
that things ``return back to normal'' in the third order of perturbative
expansion,  where  gauge parameter dependence explicitly appears.
Let us note that the ``strongest'' gauge dependence appears in 
$C_A T_R N_H$ color structures. We note in this respect that
the $N_H$-dependent divergences permit a strong check related
to decoupling of heavy quark loops from HQET. We discuss this 
issue in detail in the next Section.

\section{ HQET wave function renormalization}

There is an interesting application of the results derived above
which also serves as an additional check on its correctness.
Namely, the on-shell wave function renormalization 
constant and the known ${\overline {\rm MS}}$
wave function renormalization constant can be combined to derive 
the heavy quark wave function renormalization in HQET.
To two loops that has been done in Ref.\cite{broadhurst}.
Moreover, the HQET wave function  renormalization constant\footnote{When 
we refer to HQET wave function  renormalization constant
we mean the ${\overline {\rm MS}}$ 
wave function  renormalization constant in HQET.}
can be shown to exponentiate
, i.e. it can be written as\footnote{We are not aware of any systematic
discussion of this point within HQET. The clue that there is such 
an exponentiation, can be taken from earlier studies of 
the eikonal approximation in QCD \cite{expon}. We thank 
A.G. Grozin for discussions on this point.}
\be
Z_2^{\rm HQET}=e^{\left (\frac {\alpha_s}{\pi} C_F x_1
+\left (\frac {\alpha_s}{\pi}  \right )^2 
C_F \left (C_A x_2+T_R N_L x_3 \right )
  +\left ( \frac {\alpha_s}{\pi} \right )^3 C_F
 \left (C_A^2 x_4 + C_A T_R N_L x_5+ N_L^2 T_R^2 x_6
 +C_F N_L T_R x_7 \right ) + {\cal O}(\alpha_s^4)
\right )}.
\ee
For this reason,  only few genuinely new color structures appear 
at the three loop level and the $C_F^3$ and $C_F^2 C_A$ coefficients
 in $Z_2^{\rm HQET}$ are completely determined by 
the two loop expression for $Z_2^{\rm HQET}$.
This remarkable feature provides an additional  
strong cross check on our result.

Let us first explain why is it at all possible to determine the HQET wave
function renormalization from the ${\overline {\rm MS}}$ and the on-shell
wave function renormalization constants in full QCD. The explanation 
is apparent if one asks what HQET implies for the computation 
of Feynman diagrams.  In fact, all underlying assumptions of HQET are 
satisfied if we try to compute the quark self energy in the vicinity 
of the particle mass shell. If we denote the off-shellness
by $-\Lambda^2 = p^2-m^2$, the HQET Lagrangian gives a prescription  
of how non-analytic contributions in $\Lambda$ to the 
quark self energy should be computed. Such a computation should  
clearly be supplemented by a calculation of 
the contribution coming from  momenta region 
$k \gg \Lambda$, which is analytic in $\Lambda$. This contribution 
is provided by   the Taylor expansion of the quark self energy in 
$\Lambda$ in full QCD.   At this point the computation is reduced 
to the computation of the on-shell integrals in full QCD and this 
is precisely the piece of work we do when we compute the on-shell 
wave function  renormalization constant in full QCD 
( i.e. we compute the quark self-energy operator to the first 
non-trivial order in $\Lambda$).  The on-shell wave function
renormalization constant in full QCD is both ultraviolet and infra-red
divergent. The ultraviolet divergences can be removed by means 
of the ${\overline {\rm MS}}$ renormalization;
the infra-red divergences then cancel out once the HQET 
contribution to the quark self energy is computed since the off-shell
self energy is infra-red finite.
We therefore see, that once the infra-red divergences of the 
on-shell $Z_2$ are isolated,  they should  match the ultraviolet 
divergences in HQET and this  is the way how the HQET wave function 
renormalization constant and the HQET anomalous dimension of the heavy quark 
field are determined.

We now turn to the computation of this HQET quantity. In order to isolate
the infra-red divergences, we consider the ratio of the 
${\overline {\rm MS}}$ and the on-shell wave function renormalization 
constants in full QCD. Before giving the complete result, let us 
first discuss what we expect to happen to the contribution of 
heavy fermion  loops. 

To begin with, it is well known that the contribution of heavy 
quark loops in HQET vanishes identically due to the fact that the 
anti-particle pole is removed from the heavy quark propagator.
Therefore,  a quark-antiquark pair can not be created and 
this is the reason why the heavy fermion 
loops are disregarded  in HQET. 
Hence, it is  impossible to obtain the color structures 
proportional to $N_H$ out of any HQET computation. 
As it follows from the preceding discussion, this should imply that 
in the ratio $\displaystyle Z_2^{{\overline {\rm MS}}}/Z_2^{\rm OS}$,
computed in full QCD, any $N_H$-dependent structure should become 
finite since otherwise it would  be impossible to remove the remaining 
divergence by performing a HQET calculation.
Note, that this requirement on $N_H$-dependent structures provides a 
check on the gauge dependence of $Z_2$ computed in this paper since 
the color structure $C_A T_R N_H$ is  gauge dependent.  
By explicit computation of the ratio  
$\displaystyle Z_2^{{\overline {\rm MS}}}/Z_2^{\rm OS}$ we find, however,
that a divergent pieces proportional to $N_H$ remain in this
ratio. One should recall at this point that 
the decoupling of heavy degrees of freedom in field theories
within the minimal subtraction scheme is not automatic.
In order to observe the decoupling of heavy particles, 
one should express {\it all} $\overline {\rm MS}$ renormalized parameters 
of the theory in terms of their effective parameters whose 
renormalization scale dependence is governed by only the light degrees 
of freedom. In QCD this is done using well established decoupling relations 
\cite{dcpl1,dcpl2}.
Since the ratio $\displaystyle Z_2^{{\overline {\rm MS}}}/Z_2^{\rm OS}$
depends on the coupling constant and the gauge parameter, one should 
take into account  the decoupling relations for these two quantities in 
order to obtain a proper low energy result.
Once $\alpha_s^{(N_H+N_L)}$ and $\xi^{(N_H+N_L)}$ are expressed 
in terms of $\alpha_s^{(N_L)}$ and  $\xi^{(N_L)}$, one observes explicitly
that all divergences in the  $N_H$-depending terms disappear and one
is left with the expression that can match the structure of divergences
in HQET.  Let us emphasize that this argument alone 
proves that the on-shell wave function renormalization constant
$Z_2$ in full QCD, that we compute in this paper, should be gauge
dependent !

After this discussion, we are ready to present our result for the 
HQET anomalous dimension of the heavy quark field in the 
${\overline {\rm MS}}$ scheme. The expression for the wave function 
renormalization is then easily derived from the condition
 $\displaystyle (Z_2^{{\overline {\rm MS}}}/Z_2^{\rm OS}) Z_2^{\rm HQET} $
   = finite. We define:
\be
\frac {{\rm d} \log Z_2^{\rm HQET}}{{\rm d} \log \mu^2}
 = \gamma_{\rm HQET}.
\ee
Our result for the HQET anomalous dimension $\gamma_{\rm HQET}$ then 
reads:
\be
\gamma_{\rm HQET} = \gamma_1 \frac {\alpha_s}{\pi} + 
\gamma_2 \left ( \frac {\alpha_s}{\pi} \right )^2 +
\gamma_3 \left ( \frac {\alpha_s}{\pi} \right )^3 +
    {\cal O}(\alpha_s^4),
\ee
where
\ba
\gamma_1&& = C_F \left [ -\frac {1}{4}\xi-\frac {1}{2} \right ],
\nonumber \\
\gamma_2 && = C_F \left [ C_A \left (
-\frac {19}{24} - \frac {5}{32} \xi + \frac {1}{64}\xi^2 \right )  
+\frac {1}{3}  T_RN_L
\right ],
\nonumber \\
\gamma_3 && = C_F \Big [
C_A^2 \left ( -\frac {3}{16}\zeta_3-\frac {19495}{27648}-\frac {1}{360}\pi^4 
+ \xi \left ( -\frac {379}{2048}-\frac {15}{256}\zeta_3
+\frac {1}{1440}\pi^4  \right )                      
\right. \nonumber \\
&&  \left. + \xi^2 \left ( \frac {69}{2048}+\frac {3}{512}\zeta_3 \right )  
-\frac {5}{1024} \xi^3 \right ) + T_R^2 N_L^2 \left ( \frac {5}{108} \right )
+ C_F T_R N_L \left ( \frac {51}{64} - \frac {3}{4}\zeta_3 \right )
\nonumber \\
&& + C_A T_R N_L  \left ( \frac {1105}{6912}+\frac {3}{4}\zeta_3
+\frac {17}{256}\xi \right )
\Big ].  \label{gammaHQET}
\ea
Here $\alpha_s$ is the ${\overline {\rm MS}}$ coupling constant 
in the theory with $N_L$ massless flavors and
 the gauge parameter $\xi$ is defined in Eq.(\ref{definexi}).
The above set of equations gives the result for the ${\overline {\rm MS}}$ 
anomalous dimension of the heavy quark field in HQET up to three loops.
The two-loop result was obtained directly within HQET in \cite{brogrozin}.
The $N_L$-dependent terms in Eq.(\ref{gammaHQET}) agree with 
preliminary results of the ongoing three loop calculation 
within HQET.\footnote{
We were informed by A.Grozin that he is calculating the three-loop  HQET
anomalous dimension using a recently constructed algorithm for 3-loop
calculations in HQET \cite{grozinprogram}. He has presently obtained all
3-loop terms except for the  $C_F C_A^2$ one. The result
coincides with Eq. (\ref{gammaHQET}).}

\section{Summary of QED renormalization constants}

In this Section we summarize the results necessary for the 
on-shell renormalization of QED up to the three loop order
and also demonstrate that $Z_2$ in QED is 
gauge parameter independent in a general covariant $\xi$ gauge,
if dimensional regularization is used for both ultraviolet 
and infra-red divergences. This result for $Z_2$ was previously 
mentioned in \cite{broadhurst};
in other regularization schemes the gauge dependence of $Z_2$ was studied
in \cite{landau,johnson,fukuda}. While we do not add anything new  
to this question  in essence, we think that the derivation 
presented below is relatively simple and transparent and 
so we decided to include it into our discussion for completeness.

Let us start with deriving an equation which determines
a dependence of the fermion propagator on the gauge parameter
$\xi$ in a general covariant gauge.
Such an equation is easiest to derive in the
path integral formulation of QED. In a general
$\xi$ gauge the expression for the two-point electron Green
function reads:
\be
\langle {\rm T} \bar \psi(x) \psi(0) \rangle_\xi = N^{-1}\int [{\rm d}B]~ 
e^{\displaystyle{\frac {-i}{2\eta} \int dx B(x)^2}}
\int [d\bar \psi][d \psi] [dA] \delta( \partial_\mu A_\mu - B) \bar \psi(x)
\psi(0)~e^{\displaystyle iS},
\ee
where $\eta=1-\xi$, $S$ is the standard QED action
\be
S = \int {\rm d}^4x \left ( -\frac {1}{4} F_{\mu \nu} F^{\mu \nu} 
 + \bar \psi \left ( i \hat \partial - m \right ) \psi - e_0 \bar \psi 
 \gamma_\mu \psi A_\mu \right ),
\ee
and $N$ is the normalization factor:
\be
N = \int [{\rm d}B]~ 
e^{\displaystyle{\frac {-i}{2\eta} \int dx B(x)^2}}
\int [d\bar \psi][d \psi] [dA] \delta( \partial_\mu A_\mu - B)
~e^{\displaystyle iS}.
\ee

If the integration over 
$B$ is performed first, one recovers the
familiar expression for the path integral representation 
of the fermion two-point function  in a covariant 
gauge. We now perform a gauge transformation in the integral
over  matter fields:
\be
A_\mu \to A_\mu + \partial_\mu f,~~~~\psi \to e^{-ie_0f} \psi,
\ee
and choose $f(x)$ to satisfy the equation $\partial^2 f = B(x)$.
We  obtain:
\ba
\langle {\rm T} \bar \psi(x) \psi(0) \rangle _\xi &&
=\langle {\rm T} \bar \psi(x) \psi(0) \rangle_{B=0}
\left [ \int [{\rm d}B]~ 
e^{\displaystyle{\frac {-i}{2\eta} \int dx B(x)^2}}\right ]^{-1}
\nonumber \\
&& \times
\int [{\rm d}B(x)]~ 
\exp \left \{ {\displaystyle{\frac {-i}{2\eta} \int (dx) B^2 
+ie_0 \int (dy) \left ( G(x-y) - G(-y) \right )B(y) }}
\right \}
\ea
where $G(x)$ is the solution of the equation 
$\partial^2 G(x) = \delta^{(4)}(x)$
and $\langle {\rm T} \bar \psi(x) \psi(0) \rangle_{B=0}$ is the fermion Green
function in the Lorentz gauge $\partial_\mu A_\mu = 0$.

We can now integrate over $B(x)$ and a convenient way to 
do this is to perform a Fourier transform and integrate over its Fourier
components $B(k)$. We then obtain:
\be
\langle {\rm T} \bar \psi(x) \psi(0) \rangle_\xi
 = e^{\displaystyle \frac {i \eta e_0^2}{2} \int (dk) J_k(x)J_{-k}(x) }
\langle {\rm T} \bar \psi(x) \psi(0) \rangle_{B=0},
\label{baseq}
\ee
where $(dk) = {\rm d}^Dk/(2\pi)^D$ and $J_k$ is defined as:
\be
J_k(x) = \frac {1}{k^2} \left ( e^{ikx} - 1 \right ).
\label{jk}
\ee
By differentiating with respect to $\xi$ in Eq.(\ref{baseq}),
we obtain 
\be
\frac {{\rm d}}{{\rm d} \xi} \langle {\rm T} \bar \psi(x) \psi(0) \rangle_\xi
 =  -\frac {i e_0^2}{2} \int (dk) J_k(x)J_{-k}(x)~
 \langle {\rm T} \bar \psi(x) \psi(0) \rangle_\xi.
\ee
In momentum space this equation reads:
\be
\frac {{\rm d}}{{\rm d} \xi} S_F(p)_\xi =
 i e_0^2  \int \frac {(dq)}{q^4} S_F(p-q)_\xi
 - i e_0^2 \int \frac {(dk)}{k^4} S_F(p)_\xi.
\label{sfp}
\ee

We now analyze Eq.(\ref{sfp}) close to the mass shell where
we approximate  the  fermion propagator by:
\be
S_F(p) \approx \frac {Z_2}{\hat p - m}.
\ee
The fact that  the right hand side of Eq.(\ref{sfp}) does not develop a double
pole in the limit $p^2 \to m^2$ implies that ${\rm d}m/{\rm d}\xi$ is 
zero, in other words the pole mass of the fermion in QED is gauge 
independent.  If we equate the single pole contributions on 
both sides of Eq.(\ref{sfp}), we obtain:
\be
\frac {{\rm d}}{{\rm d} \xi} \log Z_2 
=  - i e_0^2 \int \frac {(dk)}{k^4}.
\label{final}
\ee
Note that this result is derived under the assumption that 
the integral 
\be
\int \frac {(dq)}{q^4} S_F(p-q)_\xi
\label{problem}
\ee
does not develop a pole  if $p^2 \to m^2$ and one can make sure by power
counting that this is a non-trivial statement.
The correct approach, which matches perturbative calculations,
is to take the limit $p^2 \to m^2$ first assuming that 
the integrals remain finite because of the regularization. In this 
sense the integral in Eq.(\ref{problem}) does not develop a single 
pole and can be neglected.
We now return to Eq.(\ref{final}) and note that 
in dimensional regularization the integral over $k$ is zero. Hence
we obtain:
\be
\frac {{\rm d}}{{\rm d} \xi} \log Z_2 =0.
\label{js}
\ee
We thus proved that the on-shell fermion wave function renormalization 
constant in QED is gauge parameter independent.

Let us now turn to the presentation of the three-loop QED renormalization 
constants. We consider QED of a single  electron; that implies that 
$N_H=1$ and $N_L=0$ should be used in general formulas 
of Section IV.

We write  the mass renormalization constant $Z_m$ as
\be
Z_m = 1 + 
\sum_i \left ( \frac {\alpha_0}{\pi} \frac {\Gamma(1+\ep)m^{-2\ep}}
{(4\pi)^{-\ep}} 
 \right )^i Z_m^{(i)},
\ee
where $\alpha_0$ is the bare QED coupling constant. For the expansion 
coefficients $Z_m^{(i)}$ we obtain:
\ba
Z_m^{(1)} && =-\frac {3}{4\ep}-\frac {1}{1-2\ep}, 
\nonumber \\
Z_m^{(2)} && = 
\frac {13}{32\ep^2} 
+\frac {73}{64\ep} 
+ \frac {475}{128} - \frac {3}{4}\zeta_3 + \frac {1}{2}\pi^2\log 2 
  - \frac {23}{48}\pi^2 \nonumber \\
&&
 + \ep \left ( \frac {2529}{256} - \frac {47}{4}\zeta_3 + 
 4\pi^2\log 2 - \pi^2 \log^2 2 - \frac {245}{96}\pi^2 + \frac {7}{40}\pi^4 
 - \frac {1}{2}\log^4 2 - 12 a_4 \right ),
\nonumber \\
Z_m^{(3)} && = 
 - \frac {221}{1152\ep^3} 
 - \frac {5561}{6912\ep^2} 
  +\frac {1}{\ep} \left (- \frac {154445}{41472} + \frac {13}{16}\zeta_3 
        - \frac {17}{24}\pi^2\log 2      + \frac {391}{576}\pi^2 \right )
\nonumber \\
&& 
  - \frac {3489365}{248832} - \frac {1}{16}\zeta_3\pi^2 
  + \frac {719}{72}\zeta_3 + \frac{5}{8}\zeta_5 
  + \frac {89}{36}\pi^2\log 2 + \frac {65}{36}\pi^2 \log^2 2 
\nonumber \\
&&  - \frac {5783}{17280}\pi^2 - \frac {979}{4320}\pi^4
          + \frac {23}{72}\log^4 2 + \frac {23}{3}a_4.      
\ea

The wave function renormalization constant is parameterized as
\be
Z_2 = 1 + 
\sum_i \left ( \frac {\alpha_0}{\pi (4\pi)^{-\ep} } 
\Gamma(1+\ep) \right )^i Z_2^{(i)},
\ee
where
\ba
Z_2^{(1)}&&=-\frac {3}{4\ep}-\frac {1}{1-2\ep},
\nonumber \\
Z_2^{(2)}&&= \frac {17}{32\ep^2} + \frac {229}{192\ep}
 +\frac {8453}{1152}-\frac {55}{48}\pi^2-\frac {3}{2}\zeta_3+\pi^2 \log 2 
\nonumber \\
 &&+\ep \left (\frac {31}{4}\pi^2 \log2 -\frac {419}{96}\pi^2
  +\frac {86797}{6912}-\frac {203}{8}\zeta_3-2\pi^2 \log^2 2
   -\log^42 -24 a_4+\frac {7}{20}\pi^4 \right ),
\nonumber \\
 Z_2^{(3)}&&= -\frac {131}{384\ep^3} - \frac {2141}{2304\ep^2}
                 +\frac {1}{\ep} \left (
  \frac {17}{8}\zeta_3 - \frac {17}{12}\pi^2\log2 + \frac {935}{576}\pi^2 
         - \frac {116489}{13824} \right )
\nonumber \\
 &&         -\frac {2121361}{82944}  
         + \frac {1}{8}\zeta_3 \pi^2 
          + \frac {2803}{144}\zeta_3 
         - \frac {5}{16}\zeta_5 
         + \frac {1367}{144}\pi^2 \log2 
\nonumber \\
 &&         + \frac {23}{6}\pi^2\log^2 2 - \frac {197731}{51840}\pi^2 
          - \frac {383}{720}\pi^4 
           + \frac {3}{4}\log^4 2 + 18 a_4.
\ea
We explicitly see that the  QED result for both the mass 
and the wave function  renormalization constants does not depend on the gauge 
parameter, in agreement with Johnson-Zumino identity Eq.(\ref{js})
and in variance with the QCD calculation of the previous section.

Finally, for completeness, we give the relation between the bare 
$\alpha_0$ and the physical QED coupling constant $\alpha$ 
defined in the on-shell scheme 
(i.e. through the photon propagator at zero momentum transfer). 
The result reads \cite{broadhurst}:
\be
\frac{\alpha}{\alpha_0} = 1 
- \frac {4}{3} 
\left ( \frac {\alpha_0 \Gamma(\ep)m^{-2\ep}}{\pi (4\pi)^{-\ep}} \right ) 
- \frac {4\ep}{(2-\ep)(1-2\ep)(1+2\ep)} 
 \left ( 1 + \ep (7-4\ep) \right )
\left ( \frac {\alpha_0 \Gamma(\ep)m^{-2\ep}}{\pi (4\pi)^{-\ep}} \right )^2,
\ee
and completes the set of the renormalization 
constants required for the renormalization of the QED
up to three loops in the on-shell scheme.
\section{Conclusions}

We have presented a calculation of the three loop on-shell renormalization 
constants both in QED and QCD. Explicit result for the mass and the fermion
wave function renormalization constants are derived. Dimensional 
regularization is used to regulate both ultraviolet and infrared 
divergences. 

Our calculation represents an important step towards three loop calculations 
with heavy on-shell quarks in QCD (semileptonic $b$ decays, NRQCD, etc.). 
Furthermore, these results are important for applications in QED  where 
the on-shell scheme is clearly superior as compared to any other scheme. 

The use of dimensional regularization has some interesting consequences
for the gauge dependence of the fermion wave function renormalization
constant.
First, in QED $Z_2$ turns out to be gauge invariant to all orders in 
$\alpha$.  In QCD, only the first two orders of the perturbative expansion 
of $Z_2$ in $\alpha_s$ are gauge independent; unfortunately, 
the dependence on the gauge parameter explicitly appears in 
the third order of the perturbative expansion.

As a byproduct of this analysis, we derived the anomalous dimension
for the heavy quark field in HQET.

\section*{Acknowledgments}

We are grateful to D.J. Broadhurst and A. G.~Grozin for useful conversations.
This research was supported in part by the United States
Department of Energy, contract DE-AC03-76SF00515, 
by BMBF under grant number BMBF-057KA92P, by
Gra\-duier\-ten\-kolleg ``Teil\-chen\-phy\-sik'' at the University of
Karlsruhe and by the DFG Forschergruppe 
``Quantenfeldtheorie, Computeralgebra und Monte-Carlo-Simulation''.

\newpage
\section*{Appendix A: Elementary Multiloop Integrals}
\label{sec:multloop}
In this appendix expressions are given for a number of integrals in
dimensional regularization for which exact analytic results are known.
We present these below in Minkowski space.

For a 1-loop massive bubble integral one has \cite{thooft}
\ba
\label{1loopbubble} 
\int
 && \frac{ {\rm d}^Dp\; }{(p^2-m^2+i\epsilon)^{\alpha_1}
       (p^2+i\epsilon)^{\alpha_2}} = 
\nonumber \\
 &&       i\; \pi^{D/2} (-1)^{-\alpha_1-\alpha_2}
                    (m^2)^{D/2-\alpha_1-\alpha_2}
        \frac{\Gamma(\alpha_1+\alpha_2-D/2)
          \Gamma(D/2-\alpha_2)}{\Gamma(\alpha_1)\Gamma(D/2)}.
\ea
A 1-loop massless propagator-type integral has the simple form
\ba
  \label{1loopmassless} 
&&
\int
 \frac{{\rm d}^Dp\; }{(p^2+i\epsilon)^{\alpha_1}
         [(p+Q)^2+i\epsilon]^{\alpha_2}} = 
\nonumber \\
&&     i^{1-D}\; \pi^{D/2}\; (Q^2)^{D/2-\alpha_1-\alpha_2}
       \frac{\Gamma(D/2-\alpha_1)\Gamma(D/2-\alpha_2)
         \Gamma(\alpha_1+\alpha_2-D/2)}{
        \Gamma(\alpha_1)\Gamma(\alpha_2)\Gamma(D-\alpha_1-\alpha_2)}.
\ea
A compact expression for this integral with a general tensor numerator
can be found in Ref. \cite{ibp}.
For a 1-loop on-shell propagator-type integral one derives
\ba
       \label{1looponshell} 
&&
\int
 \frac{{\rm d}^Dp\; }{(p^2+i\epsilon)^{\alpha_1}
      (p^2+2p\cdot Q+i\epsilon)^{\alpha_2}} = 
\nonumber \\
 &&  i\; \pi^{D/2}(Q^2)^{D/2-\alpha_1-\alpha_2} (-1)^{-\alpha_1-\alpha_2}
       \frac{\Gamma(\alpha_1+\alpha_2-D/2)\Gamma(D-2\alpha_1-\alpha_2)}{
        \Gamma(\alpha_2)\Gamma(D-\alpha_1-\alpha_2)},
\ea
and for a two-loop bubble integral with one massless and two massive lines
one finds \cite{BijVelt}
\ba
&& \int\int \frac{{\rm d}^Dp\; {\rm d}^Dk\;}{(p^2-m^2+i\epsilon)^{\alpha_1}\;
     [(p+k)^2+i\epsilon]^{\alpha_2}\;(k^2-m^2+i\epsilon)^{\alpha_3}} = 
\nonumber \\
&&
    \pi^D  (m^2)^{D-\alpha_1-\alpha_2-\alpha_3} \;
                (-1)^{1 -\alpha_1-\alpha_2-\alpha_3}
           \frac{\Gamma(-D+\alpha_1+\alpha_2+\alpha_3)}{
               \Gamma(\alpha_1)\Gamma(\alpha_3)}
\nonumber  \\
&&   \times \frac{ \Gamma(-D/2+\alpha_1+\alpha_2)\Gamma(-D/2+\alpha_2+\alpha_3)
       \Gamma(D/2-\alpha_2)}{\Gamma(D/2)
       \Gamma(\alpha_1+2\alpha_2+\alpha_3-D)}.  \hspace{2cm}
\label{bubblem0m}
\ea
Also for this integral with a general tensor numerator
a compact expression is known \cite{2loopbubclosed}.

Several more simple cases follow by using these expressions recursively
as the powers $\alpha_1$,$\alpha_2$ and  $\alpha_3$, are allowed to be
non-integer, possibly containing $D$.
In this way one can obtain, for instance,
\ba
&& \int\int
 \frac{{\rm d}^Dp\; {\rm d}^Dk }{(p^2+i\epsilon)^{\alpha_1}\;
      (k^2-m^2+i\epsilon)^{\alpha_2}\;
       (k^2+i\epsilon)^{\alpha_4}\; [(p+k)^2+i\epsilon]^{\alpha_3}} = 
\nonumber \\
&&     \pi^D  (m^2)^{D-\alpha_1-\alpha_2-\alpha_3-\alpha_4} \;
                (-1)^{1 -\alpha_1-\alpha_2-\alpha_3-\alpha_4}
      \frac{\Gamma(\alpha_1+\alpha_2+\alpha_3+\alpha_4 -D)}{
          \Gamma(\alpha_1)\Gamma(\alpha_2)\Gamma(\alpha_3)} 
\nonumber \\
&&       \times  \frac{
     \Gamma(\alpha_1+\alpha_3-D/2)\Gamma(D/2-\alpha_1)
       \Gamma(D/2-\alpha_3)\Gamma(D-\alpha_1-\alpha_3-\alpha_4)}{
 \Gamma(D/2) \Gamma(D-\alpha_1-\alpha_3)}.
\label{bubblem00}
\ea

\section*{Appendix B: some harmonic sums}

In this Appendix we present a partial list of harmonic sums 
which we used in this paper. These results
can be found e.g.  in  
\cite{infinitesums1,infinitesums2,infinitesums3,sumsjos}.

\begin{eqnarray} \displaystyle 
\label{sumlistbegin}
 \sum_{i=1}^{\infty} \frac{1}{i^k} & = & S_k(\infty)
        = \zeta_k  \hspace{1cm} (k > 1)  \\ \displaystyle
 \sum_{i=1}^{\infty} \frac{S_1(i)}{i^2} & \equiv & S_{2,1}(\infty) = 2 \zeta_3,
 \\ \displaystyle
 \sum_{i=1}^{\infty} \frac{S_1(i)}{i^3}
    &  \equiv  & \displaystyle  S_{3,1}(\infty) = \frac{5}{4}\zeta_4,
\\ \displaystyle
\sum_{i=1}^{\infty} \frac{S_1(i)}{i^4}
    &  \equiv  & \displaystyle  S_{4,1}(\infty) = 3\zeta_5 - \zeta_2\zeta_3, 
\\ \displaystyle
 \sum_{i=1}^{\infty} \frac{S_2(i)}{i^2} & \equiv & S_{2,2}(\infty) =
        \displaystyle \frac{7}{4} \zeta_4
,\\ \displaystyle
 \sum_{i=1}^{\infty} \frac{S_2(i)}{i^3} & \equiv & S_{3,2}(\infty) =
    \displaystyle   3\zeta_2\zeta_3 - \frac{9}{2}\zeta_5
,\\ \displaystyle
\sum_{i=1}^{\infty} \frac{ S_{2,1}(i)}{i^2} &  \equiv & S_{2,2,1}(\infty)
         \displaystyle  =  2 \zeta_5
,\\ \displaystyle 
 \sum_{i=1}^{\infty} \frac{ S^2_1(i)}{i^2}  & = &
         \displaystyle     \frac{17}{4}\zeta_4
,\\ \displaystyle 
 \sum_{i=1}^{\infty} \frac{ S^3_1(i)}{i^2}  & = &
         \displaystyle     \zeta_2\zeta_3 + 10\zeta_5
,\\ \displaystyle 
 \sum_{i=1}^{\infty} \frac{ S^2_1(i)}{i^3}  & = &
         \displaystyle    \frac{7}{2}\zeta_5    - \zeta_2\zeta_3
,\\ \displaystyle
 \sum_{i=1}^{\infty} \frac{ S_3(i)}{i^2}  & = &
         \displaystyle   \frac{11}{2}\zeta_5  - 2 \zeta_2\zeta_3
,\\ \displaystyle 
 \sum_{i=1}^{\infty} \frac{ S_1(i)S_2(i)}{i^2}  & = &
         \displaystyle   \zeta_5 + \zeta_2\zeta_3. 
\label{sumlistend}
\end{eqnarray}

\section*{Appendix C: List of master integrals}

In this Appendix we present the list of master integrals we have used 
in our calculation. The list updates the result that can be found in 
\cite{Laporta}.  The Euclidean master integrals are defined as (for a pictorial
representation see Fig.2):
\ba
&&I_1 = 
\left \langle \frac {(-1)~p \cdot k_2}{D_1 D_2 D_3 D_4 D_5 D_6 D_7 D_8} 
 \right \rangle,
\nonumber \\
&&I_2 = \left   \langle \frac {1}{D_1 D_2 D_3 D_4 D_7 D_8} \right \rangle,
~~~~~~~~~~~~~~
I_3 =  \left \langle \frac {1}{D_1 D_2 D_4 D_5 D_6 D_8} \right \rangle,
\nonumber \\
&&I_4 =  \left \langle \frac {1}{D_2 D_3 D_4 D_6 D_7 D_8} \right \rangle,
~~~~~~~~~~~~~~
I_5 =  \left \langle \frac {1}{D_1 D_3 D_4 D_5 D_7 D_8} \right \rangle,
\nonumber \\
&&I_6 = \left \langle \frac {1}{D_1 D_3 D_5 D_6 D_7 D_8} \right \rangle,
~~~~~~~~~~~~~~
I_7 = \left \langle \frac {1}{D_2 D_4 D_5 D_6 D_7 D_8} \right \rangle,
\nonumber \\
&&I_8 = \left \langle \frac {1}{D_1 D_2 D_3 D_4 D_5} \right \rangle,
~~~~~~~~~~~~~~~~~~
I_9 = \left \langle \frac {1}{D_2 D_3 D_5 D_6 D_7} \right \rangle,
\nonumber \\
&&I_{10} = \left \langle \frac {1}{D_2 D_4 D_6 D_7 D_8} \right \rangle,
~~~~~~~~~~~~~~~~~
I_{11} = \left \langle \frac {1}{D_1 D_3 D_5 D_7} \right \rangle,
\nonumber \\
&&I_{12} = \left \langle \frac {1}{D_1 D_2 D_4 D_5} \right \rangle,
~~~~~~~~~~~~~~~~~~~~~
I_{13} = \left \langle \frac {1}{D_3 D_5 D_6 D_7} \right \rangle,
\nonumber \\
&&I_{14} = \left \langle \frac {1}{D_2 D_3 D_4 D_5} \right \rangle,
~~~~~~~~~~~~~~~~~~~~~
I_{15} = \left \langle \frac {1}{D_3 D_4 D_7 D_8} \right \rangle,
\nonumber \\
&&I_{16} = \left \langle \frac {1}{D_3 D_6 D_7 D_8} \right \rangle,
~~~~~~~~~~~~~~~~~~~~~
I_{17} = \left \langle \frac {1}{D_1 D_4 D_5} \right \rangle,
\nonumber \\
&&I_{18} =\left \langle \frac {1}{D_1 D_2 D_5 D_6 D_7 D_8 D_9} \right \rangle,
\ea
where 
\be
\left \langle ... \right \rangle  = \int {\rm d}^D k_1  {\rm d}^D k_2
 {\rm d}^D k_3 ......, 
\ee
the momentum $p$ is always considered as {\it incoming} and 
\ba
&& D_1 = (p-k_1)^2 + 1,~~~~~~~~~~~~~~~~~~~~~~~~D_2 = (p-k_1-k_2)^2 +1,
\nonumber \\
&& D_3 = (p-k_1-k_2-k_3 )^2 + 1,~~~~~~~~~~~D_4 = (p-k_2-k_3)^2 +1,
\nonumber \\
&& D_5 = (p-k_3)^2 + 1,~~~~~~~~~~~~~~~~~~~~~~~~D_6 = k_1^2,
\nonumber \\
&& D_7 = k_2^2,~~~~~~~~~~~~~~~~~~~~~~~~~~~~~~~~~~~~~~D_8 = k_3^2,
\nonumber \\
&& D_9 = (k_3-k_1-k_2)^2.
\ea
where $p^2 = -1$.

\ba
I_1 &&= C(\ep) \left (-5 \zeta_5 + \frac {1}{2} \pi^2 \zeta_3 + {\cal O}(\ep) 
 \right ),
\nonumber \\
I_2 &&= C(\ep) \left (
      2 \frac {\zeta_3}{\ep} 
       + 10\zeta_3 - \frac {1}{3}\pi^2 - \frac {13}{90}\pi^4
     + \ep \left (\frac {49}{6}\pi^2 \zeta_3 + 24\zeta_3 - \frac {85}{2}\zeta_5        + 4\pi^2\log 2 - \frac {4}{3}\pi^2 - \frac {13}{18}\pi^4  \right ) 
 + {\cal O}(\ep^2)
 \right ),
\nonumber \\
I_3 &&= C(\ep) \left ( 
           \frac {1}{3\ep^3} 
          + \frac {7}{3\ep^2} 
          + \frac {31}{3\ep} 
       + \frac {103}{3} - \frac {4}{3}\zeta_3 - \frac {2}{15}\pi^4
    + \ep \left ( 
 \frac {235}{3} + \frac {28}{3}\zeta_3\pi^2 + \frac {32}{3}\zeta_3 
     - 78\zeta_5 + \frac {8}{3}\pi^2 - \frac {3}{5}\pi^4  \right )
 + {\cal O}(\ep^2)
\right ),
\nonumber \\
I_4&&= C(\ep) \left ( 2 \frac {\zeta_3}{\ep}
       + 2\zeta_3 + \frac {1}{3}\pi^2 - \frac {7}{90}\pi^4
       + \ep \left (  
 - \frac {2}{3}\zeta_3\pi^2 - 12\zeta_3 + 44\zeta_5 + \frac {14}{3}\pi^2 
  - \frac {41}{90}\pi^4  \right )  + {\cal O}(\ep^2)
\right ),
\nonumber \\
I_5 &&=    C(\ep) \left (
      \frac {1}{6\ep^3} 
      +\frac {3}{2\ep^2}
       + \frac {1}{\ep} \left ( \frac {55}{6} - \frac {1}{3}\pi^2 \right )
       + \frac {95}{2} - \frac {14}{3}\zeta_3 - \frac {7}{3}\pi^2 
               - \frac {4}{45}\pi^4
\right. \nonumber \\
&& \left.       + \ep \left ( \frac {1351}{6} + \frac {25}{6}\zeta_3\pi^2 
       - 42\zeta_3 - \frac {49}{2}\zeta_5 + 4\pi^2\log 2 
          - \frac {31}{3}\pi^2 - \frac {23}{30}\pi^4  \right ) 
 + {\cal O}(\ep^2) \right ),
\nonumber \\
I_6&& = C(\ep) \left (
       \frac {1}{3\ep^3} 
       + \frac {7}{3\ep^2} 
       +\frac {31}{3\ep} 
       + \frac {103}{3} + \frac {2}{3}\zeta_3 + \frac {1}{3}\pi^2 
             - \frac {4}{45}\pi^4
       + \ep  \left ( 
    \frac {235}{3} + \frac {2}{3}\zeta_3 \pi^2 
    + \frac {20}{3}\zeta_3 - 2\zeta_5 + 4\pi^2 - \frac {3}{10}\pi^4 
          \right  ) + {\cal O}(\ep^2)
\right ),
\nonumber \\
I_7&& =  C(\ep) \left (
       \frac {1}{6\ep^3} 
       + \frac {3}{2\ep^2} 
       + \frac {1}{\ep}  \left ( \frac {55}{6} - \frac {1}{3}\pi^2 \right )
       + \frac {95}{2} - \frac {8}{3}\zeta_3 - 2\pi^2 - \frac {1}{15}\pi^4
\right. \nonumber \\
&& \left.
       + \ep \left ( 
  \frac {1351}{6} + 6\zeta_3\pi^2 - 14\zeta_3 
 - 64\zeta_5 - \frac {17}{3}\pi^2 - \frac {47}{45}\pi^4 \right )
 + {\cal O}(\ep^2)
\right ),
\nonumber \\
I_8 &&=  C(\ep) \left (
       - \frac {1}{\ep^3} 
       - \frac {16}{3\ep^2}
       - \frac {16}{\ep} 
       - 20 + 2\zeta_3 - \frac {8}{3}\pi^2
       + \ep  \left ( \frac {364}{3} 
     - \frac {200}{3}\zeta_3 + 16\pi^2\log 2 - 28\pi^2 - \frac {3}{10}\pi^4 
          \right )
\right. \nonumber \\
&& \left.
       + \ep^2  \left ( 
       1244 + 21\zeta_3\pi^2 - 776\zeta_3 - 126\zeta_5 + 168\pi^2\log 2 
         - \frac {80}{3}\pi^2\log^2 2 - 188\pi^2 
   + \frac {46}{15}\pi^4 - \frac {64}{3}\log^4 2 - 512 a_4 \right )
 + {\cal O}(\ep^3)
\right ),
\nonumber \\
I_9 &&=  C(\ep) \left (
       - \frac {2}{3\ep^3} 
       - \frac {10}{3\ep^2} 
       + \frac {1}{\ep} \left (  - \frac {26}{3} - \frac {1}{3}\pi^2  \right )
       - 2 - \frac {16}{3}\zeta_3 - \frac {11}{3}\pi^2
      +\ep  \left ( 
  \frac {398}{3} - \frac {248}{3}\zeta_3 
 + 16\pi^2\log 2 - \frac {73}{3}\pi^2 - \frac {13}{45}\pi^4 \right )
\right. \nonumber \\
&& \left.
       + \ep^2 \left ( 
   1038 - \frac {8}{3}\zeta_3 \pi^2 - \frac {1888}{3}\zeta_3 
      - 96\zeta_5 + 160\pi^2\log 2 - 
         \frac {128}{3}\pi^2\log^2 2 - 129\pi^2 + \frac {3}{5}\pi^4 
       - \frac {64}{3}\log^4 2 - 512 a_4 \right ) + {\cal O}(\ep^3)
\right ),
\nonumber \\
I_{10} && =  C(\ep) \left (
       - \frac {1}{3\ep^3}
       - \frac {5}{3\ep^2} 
       + \frac {1}{\ep} \left (  - 4 - \frac {2}{3}\pi^2 \right )
       + \frac {10}{3} - \frac {26}{3}\zeta_3 - \frac {7}{3}\pi^2
       + \ep \left ( \frac {302}{3} - \frac {94}{3}\zeta_3 
           - \pi^2 - \frac {35}{18}\pi^4 \right )
\right. \nonumber \\
&& \left.
       - \ep^2  \left (  -734 +\frac {76}{3} \pi^2\zeta_3 
       -\frac {101}{3}\pi^2 
      +20\zeta_3 +\frac {551}{90}\pi^4 +462\zeta_5 \right ) + {\cal O}(\ep^3)
\right ),
\nonumber \\
I_{11} && =  C(\ep) \left (
         \frac {1}{\ep^3} 
       + \frac {7}{2\ep^2} 
       +  \frac {253}{36\ep} 
       + \frac {2501}{216}
       + \ep  \left  ( \frac {59437}{1296} - \frac {64}{9}\pi^2 \right )
       + \ep^2 \left( 
 \frac {2831381}{7776} - \frac {1792}{9}\zeta_3 
 + \frac {256}{3}\pi^2 \log 2 - \frac {2272}{27}\pi^2  \right )
\right. \nonumber \\
&& \left.
       + \ep^3  \left ( 
 \frac {117529021}{46656} - \frac {63616}{27}\zeta_3 
 + \frac {9088}{9}\pi^2 \log 2 - \frac {3584}{9}\pi^2 \log^2 2 
      - \frac {49840}{81}\pi^2 + \frac {2752}{135}\pi^4 
 - \frac {1024}{9}\log^4 2 - \frac {8192}{3} a_4  \right )
+ {\cal O}(\ep^4)
 \right ),
\nonumber \\
I_{12}&& =  C(\ep) \left (
       \frac {2}{\ep^3} 
      + \frac {23}{3\ep^2} 
       + \frac {35}{2\ep} 
       + \frac {275}{12}
       + \ep  \left (  - \frac {189}{8} + \frac {112}{3} \zeta_3 \right )
\right. \nonumber \\
&& \left.
       + \ep^2 \left (  
 - \frac {14917}{48} + 280\zeta_3 - \frac {32}{3}\pi^2\log^2 2 
 - \frac {136}{45}\pi^4 
       + \frac {32}{3}\log^4 2 + 256 a_4 \right )
      + {\cal O}(\ep^3) 
\right),
\nonumber \\
I_{13}&& =  C(\ep) \left (
       \frac {1}{3\ep^3} 
       + \frac {7}{6\ep^2} 
       + \frac {25}{12\ep} 
       - \frac {5}{24} + \frac {8}{3}\zeta_3
       + \ep \left (  
 - \frac {959}{48} + \frac {28}{3}\zeta_3 - \frac {2}{15}\pi^4 
             \right  )
\right. \nonumber \\
&& \left.
       + \ep^2  \left (  - \frac {10493}{96} + \frac {50}{3}\zeta_3 
     + 48\zeta_5 - \frac {7}{15}\pi^4 \right )
     + {\cal O}(\ep^3) 
\right ),
\nonumber \\
I_{14}&& =  C(\ep) \left (
       \frac {3}{2\ep^3} 
       + \frac {23}{4\ep^2}
       + \frac {105}{8\ep} 
       + \frac {275}{16} + \frac {4}{3}\pi^2
       + \ep  \left (  - \frac {567}{32} + 28\zeta_3 - 8\pi^2\log 2 
     + 10\pi^2 \right )
\right. \nonumber \\
&& \left.
       + \ep^2  \left (  
    - \frac {14917}{64} + 210\zeta_3 - 60\pi^2\log 2 + 16\pi^2\log^2 2 
      + \frac {145}{3}\pi^2 - \frac {62}{45}\pi^4 + 8\log^4 2 + 192a_4 
        \right )
     + {\cal O}(\ep^3) 
\right ),
\nonumber \\
I_{15}&& = C(\ep) \left (
        \frac {1}{2\ep^3} 
       + \frac {7}{4\ep} 
       + \frac {1}{\ep}  \left ( \frac {25}{8} + \frac {1}{3}\pi^2 \right )
       - \frac {5}{16} + 4\zeta_3 + \frac {7}{6}\pi^2
       + \ep  \left (  - \frac {959}{32} + 14\zeta_3 + \frac {25}{12}\pi^2 
        + \frac {16}{45}\pi^4 \right )
\right. \nonumber \\
&& \left.
       + \ep^2  \left (  - \frac {10493}{64} + \frac {8}{3}\zeta_3\pi^2 
    + 25\zeta_3 + 72\zeta_5 - \frac {5}{24}\pi^2 
        + \frac {56}{45} \pi^4 \right ) 
    + {\cal O}(\ep^3) 
\right ),
\nonumber \\
I_{16}&& = C(\ep) \left (
       - \frac {1}{6\ep^3} 
       - \frac {5}{6\ep^2} 
       + \frac {1}{\ep} \left (  - \frac {11}{6} - \frac {1}{3}\pi^2 \right )
       + \frac {23}{6} - \frac {16}{3}\zeta_3 - \frac {5}{3}\pi^2 
      + \ep \left ( \frac {135}{2} - \frac {80}{3}\zeta_3 
      - \frac {11}{3}\pi^2 - \frac {37}{45}\pi^4 \right )
\right. \nonumber \\
&& \left.
       + \ep^2 \left ( \frac {949}{2} - \frac {32}{3}\pi^2\zeta_3 
      - \frac {176}{3}\zeta_3 - 208\zeta_5 + \frac {23}{3}\pi^2 
         - \frac {37}{9} \pi^4 \right ) + {\cal O}(\ep^3)
\right ),
\nonumber \\
I_{17}&& = C(\ep) \left (
       - \frac {1}{\ep^3} 
       - \frac {3}{\ep^2} 
       - \frac {6}{\ep} 
       -10
       - 15\ep 
        - 21\ep^2 
         - 28\ep^3  + {\cal O}(\ep^4) \right),
\nonumber \\
I_{18}&& = C(\ep) \left ( 2\pi^2\zeta_3 - 5\zeta_5 
          + {\cal O}(\ep) \right).
\nonumber
\ea

 Finally, we should note that integrals $I_{12}$ and $I_{14}$ are known 
 to a higher order in $\varepsilon$ than we have written here;
 the corresponding higher order terms in $\varepsilon$ can be found in 
  Ref. \cite{broadhurstprimitives}. Furthermore,
 integrals $I_{13}$, $I_{15}$, $I_{16}$ and $I_{17}$ can be 
  performed exactly in terms of Euler gamma functions, as one can see
  for instance by combining the exact forms in Appendix A. This means that
  these four integrals are known in principle to an arbitrary high order in 
  $\varepsilon$.

\newpage
\begin{figure}
\begin{center}
\hfill
\begin{picture}(480,160)(0,160)

\SetScale{.5}

\SetOffset(0,300)

 \SetWidth{3}
 \Line(0,25)(10,25)
 \Line(10,25)(30,50)
 \Line(30,50)(70,50)
 \Line(70,50)(30,0)
 \Line(30,0)(70,0)
 \Line(70,0)(90,25)
 \Line(90,25)(100,25)
 \SetWidth{.5}
 \Line(10,25)(30,0)
 \Line(30,50)(47,28)
 \Line(53,22)(70,0)
 \CArc(50,25)(5,135,315)
 \Line(90,25)(70,50)
\Text(5,0)[m]{$I_1$}

\SetOffset(80,300)

 \SetWidth{3}
 \Line(0,25)(10,25)
 \Line(50,50)(90,25)
 \Line(50,50)(50,0)
 \Line(50,0)(90,25)
 \Line(90,25)(100,25)
 \CArc(50,22)(40,5,175)
 \SetWidth{.5}
 \Line(10,25)(50,50)
 \Line(10,25)(50,0)
 \Text(5,0)[m]{$I_2$}

\SetOffset(160,300)

 \SetWidth{3}
 \Line(0,25)(10,25)
 \Line(10,25)(50,50)
 \Line(50,50)(90,25)
 \CArc(24,25)(36,318,42)
 \CArc(76,25)(36,138,222)
 \Line(90,25)(100,25)
 \SetWidth{.5}
 \Line(10,25)(50,0)
 \Line(50,0)(90,25)
 \Text(5,0)[m]{$I_3$}

\SetOffset(240,300)

 \SetWidth{3}
 \Line(0,25)(10,25)
 \Line(10,25)(50,50)
 \Line(50,50)(50,0)
 \Line(50,0)(90,25)
 \Line(90,25)(100,25)
 \SetWidth{.5}
 \Line(10,25)(50,0)
 \Line(50,50)(90,25)
 \CArc(50,22)(40,5,175)
 \Text(5,0)[m]{$I_4$}

\SetOffset(320,300)
 \SetWidth{3}
 \Line(0,25)(10,25)
 \Line(10,25)(50,50)
 \Line(50,50)(90,25)
 \CArc(49,45)(45,271,334)
 \CArc(90,-19.5)(45,92,154)
 \Line(90,25)(100,25)
 \SetWidth{.5}
 \Line(10,25)(50,0)
 \Line(50,0)(50,50)
 \Text(5,0)[m]{$I_5$}

\SetOffset(400,300)
 \SetWidth{3}
 \Line(0,25)(10,25)
 \Line(10,25)(50,50)
 \CArc(24,25)(36,318,42)
 \Line(50,0)(90,25)
 \Line(90,25)(100,25)
 \SetWidth{.5}
 \Line(10,25)(50,0)
 \Line(50,50)(90,25)
 \CArc(76,25)(36,138,222)
 \Text(5,0)[m]{$I_6$}

\SetOffset(0,240)
 \SetWidth{3}
 \Line(0,25)(10,25)
 \Line(10,25)(50,50)
 \Line(50,0)(50,50)
 \CArc(49,45)(45,271,334)
 \Line(90,25)(100,25)
 \SetWidth{.5}
 \Line(10,25)(50,0)
 \Line(50,50)(90,25)
 \CArc(90,-19.5)(45,92,154)
 \Text(5,0)[m]{$I_7$}

\SetOffset(80,240)
 \SetWidth{3}
 \Line(0,25)(10,25)
 \CArc(50,45)(45,206,334)
 \CArc(90,-19.5)(45,92,154)
 \CArc(10,-19.5)(45,26,89)
 \CArc(50,0)(48,30,150)
 \Line(90,25)(100,25)
 \SetWidth{.5}
 \Text(5,0)[m]{$I_8$}

\SetOffset(160,240)
 \SetWidth{3}
 \Line(0,25)(10,25)
 \CArc(50,45)(45,206,270)
 \CArc(10,-19.5)(45,26,89)
 \CArc(50,0)(48,30,150)
 \Line(90,25)(100,25)
 \SetWidth{.5}
 \CArc(50,45)(45,270,334)
 \CArc(90,-19.5)(45,92,154)
 \Text(5,0)[m]{$I_9$}

\SetOffset(240,240)
 \SetWidth{3}
 \Line(0,25)(10,25)
 \CArc(50,45)(45,206,334)
 \Line(90,25)(100,25)
 \SetWidth{.5}
 \CArc(90,-19.5)(45,92,154)
 \CArc(10,-19.5)(45,26,89)
 \CArc(50,0)(48,30,150)
 \Text(5,0)[m]{$I_{10}$}

\SetOffset(320,240)
 \SetWidth{3}
 \Line(0,25)(10,25)
 \CArc(50,45)(45,206,334)
 \CArc(50,-45)(80,60,120)
 \CArc(50,95)(80,240,300)
 \Line(90,25)(100,25)
 \SetWidth{.5}
 \CArc(50,5)(45,25,155)
 \Text(5,0)[m]{$I_{11}$}

\SetOffset(400,240)
 \SetWidth{3}
 \Oval(55,25)(25,30)(0)
 \CArc(29,25)(36,318,42)
 \CArc(81,25)(36,138,222)
 \Line(30,-2)(80,-2)
 \SetWidth{.5}
 \Text(5,0)[m]{$I_{12}$}

\SetOffset(0,180)
 \SetWidth{3}
 \Oval(55,25)(25,30)(0)
 \Line(30,-2)(80,-2)
 \SetWidth{.5}
 \CArc(29,25)(36,318,42)
 \CArc(81,25)(36,138,222)
 \Text(5,0)[m]{$I_{13}$}

\SetOffset(80,180)
 \SetWidth{3}
 \Line(-7,24)(5,24)
 \CArc(45,7)(45,23,157)
 \CArc(45,-43)(78,57,123)
 \CArc(45,167)(150,254,286)
 \Oval(89,7.8)(14,12)(0)
 \Line(85,24)(103,24)
 \SetWidth{.5}
 \Text(5,0)[m]{$I_{14}$}

\SetOffset(160,180)
 \SetWidth{3}
 \Line(-7,24)(5,24)     
 \CArc(45,167)(150,254,286)   
 \Oval(89,7.8)(14,12)(0)
 \Line(85,24)(103,24)     
 \SetWidth{.5}
 \CArc(45,7)(45,23,157)
 \CArc(45,-43)(78,57,123)  
 \Text(5,0)[m]{$I_{15}$}

\SetOffset(240,180)
 \SetWidth{3}
 \Line(0,25)(10,25)
 \CArc(50,45)(45,206,334)
 \Line(90,25)(100,25)
 \SetWidth{.5}
 \CArc(50,5)(45,25,155)
 \CArc(50,-45)(80,60,120)
 \CArc(50,95)(80,240,300)
 \Text(5,0)[m]{$I_{16}$}

\SetOffset(320,180)
 \SetWidth{3}
 \Line(20,23)(80,23)
 \CArc(50,65)(40,225,315)
 \CArc(68,46)(14,315,90) 
 \CArc(68.5,48)(12,90,158) 
 \CArc(125.6,22)(75,155,177)
 \CArc(-25.6,22)(75,3,25)
 \CArc(32,46)(14,90,225)
 \CArc(31.5,48)(12,22,90)
 \Oval(50,5.5)(16,14)(0)
 \SetWidth{.5}
 \Text(5,0)[m]{$I_{17}$}

\SetOffset(400,180)

 \SetWidth{3}
 \Line(0,25)(10,25)
 \CArc(50,45)(45,206,334)
 \Line(90,25)(100,25)
 \SetWidth{.5}
 \CArc(50,0)(48,30,150)
 \Line(50,48)(70,5)
 \Line(50,48)(30,5)
 \Text(5,0)[m]{$I_{18}$}

\end{picture}
\end{center}
\caption{Pictorial representation of the complete set of primitive 3-loop 
  on-shell integrals $I_1$ -- $I_{18}$. Thick lines are used to 
  indicate massive scalar propagators and thin lines indicate massless 
   propagators.}
\end{figure}
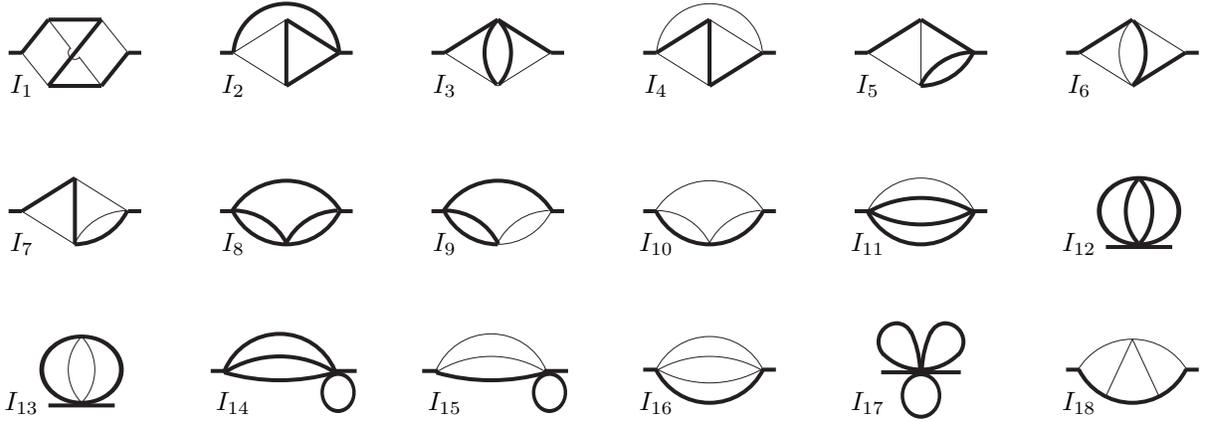


\end{document}